\documentclass[aps,amssymb,amsmath,amsfonts,twocolumn]{revtex4}
\usepackage{epsfig}
\usepackage{bm,bbm}
\usepackage{verbatim} 

\pdfoutput=1

\newcommand{\bR}{{\mathbb R}}
\newcommand{\id}{{\rm 1\mskip-4mu l}}

\begin{document}
\title{Geometry of the set of mixed quantum states: An apophatic approach}

\medskip

\author{Ingemar Bengtsson$^1$, Stephan Weis$^2$ and Karol \.Zyczkowski$^{3,4}$} 

\affiliation{${}^1$Stockholms Universitet, Fysikum, Roslagstullsbacken 21,
  106\,91 Stockholm, Sweden}

\affiliation{${}^2$Max Planck Institute for Mathematics in the Sciences,
Inselstrasse 22, 04103 Leipzig, Germany}

\affiliation{${}^3$Institute of Physics,
    Jagiellonian University, ul. Reymonta 4, 30-059 Krak{\'o}w, Poland}

\affiliation{${}^4$Center for Theoretical Physics, Polish Academy of Sciences, Aleja Lotnik{\'o}w 32/46, PL-02-668 Warsaw, Poland}


\medskip

\date{{December 10, 2011}}

\begin{abstract}
The set of quantum states consists of density matrices of order $N$,
which are hermitian, positive and normalized by the trace condition.
We analyze the structure of this set in the framework of the
Euclidean geometry naturally arising in the space of hermitian 
matrices. For $N=2$ this set is the Bloch ball, embedded in
$\mathbbm R^3$. For $N \ge 3$ this set of dimensionality $N^2-1$
has a much richer structure. We study its properties and at first
advocate an apophatic approach, which concentrates on characteristics
not possessed by this set. We also apply more constructive techniques
and analyze two dimensional cross-sections and projections 
of the set of quantum states. They are dual to each other.
At the end we make some remarks on certain dimension dependent 
properties.
\end{abstract}

\maketitle

{\sl Dedicated to prof. Bogdan Mielnik on the occasion of his 75-th birthday} 

\section{Introduction}

Quantum information processing differs significantly
from processing of classical information.
This is  due to the fact that the space of all states
allowed in the quantum theory is much richer
than the space of classical states \cite{Mi68,ACH93,MW98,AS03,GKM05,BZ06}.
Thus an author of a quantum algorithm, writing
a screenplay designed specially for the quantum scene,
can rely on states and transformations not 
admitted by the classical theory.

For instance, in the theory of classical information 
the standard operation of inversion of a bit, 
called the {\sl NOT} gate, cannot be represented as a concatenation
of two identical operations on a bit. But 
the quantum theory
allows one to construct the gate called $\sqrt{\rm NOT}$, 
which performed twice is equivalent to the flip of a qubit.

This simple example can be explained by comparing the
geometries of classical and quantum state spaces. 
Consider a system containing $N$ perfectly distinguishable states.
In the classical case the set of classical states, 
equivalent to $N$--point probability distributions,
forms a regular simplex $\Delta_{N-1}$ in $N-1$ dimensions.
Hence the set of pure classical states consists of $N$ isolated points.
In a quantum set-up the set of states ${\cal Q}_N$, consisting of hermitian,
positive and normalized density matrices, has $N^2-1$ real dimensions.
Furthermore, the set of pure quantum states is connected,
and for any two pure states there exist transformations that take us 
along a continuous path joining the two quantum pure states.
This fact is one of the key differences between the classical
and the quantum theories \cite{Ha01}.

The main goal of the present work is to provide 
an easy-to-read description of similarities and differences
between the sets of classical and quantum states. 
Already when $N=3$ the geometric
structure of the eight dimensional set ${\cal Q}_3$ is not easy to
analyse nor to describe \cite{Bl76,AMM97}. Therefore 
we are going to use an {\sl apophatic approach},
in which one tries to describe the properties of a given object
by specifying simple features it {\sl does not} have.
Then we use a more conventional \cite{JS01,VDM02,Sollid} constructive 
approach and investigate two-dimensional cross-sections
and projections of the set ${\cal Q}_3$ \cite{Weis_tc,Weis_cs,Dunkl}. 
Thereby a cross-section is defined as the intersection of a given set
with an affine space. We happily recommend a very recent work for a
more exhaustive discussion of the cross-sections \cite{Goyal}.

\section{Classical and quantum states}

A classical state is a probability vector $\vec{p} =(p_1,p_2,\dots, p_N)$, 
such that $\sum_i p_i=1$ and $p_i \ge 0$ for $i=1,\dots N$. Assuming that 
a pure quantum state $|\psi\rangle$ 
belongs to an $N$--dimensional Hilbert space ${\cal H}_N$, a general quantum 
state is a density matrix $\rho$ of size $N$, which is hermitian, 
$\rho=\rho^{\dagger}$, with positive eigenvalues, $\rho \ge 0$, and 
normalized,  Tr$\rho=1$. Note that any density matrix can be diagonalised, 
and then it has a probability vector along its diagonal. But clearly 
the space of all quantum states ${\cal Q}_N$ is significantly larger 
than the space of all classical states---there are $N-1$ free parameters 
in the probability vector, but there are $N^2-1$ free parameters in the 
density matrix. 

The space of states, classical or quantum, is always a {\sl convex} set. 
By definition a convex set is a subset of Euclidean space, such that 
given any two points in the subset the line segment between the two 
points also belongs to that subset. The points in the interior of the 
line segment are said to be {\sl mixtures} of the original points. Points 
that cannot be written as mixtures of two distinct points are called 
{\sl extremal} or {\sl pure}. Taking all mixtures of three pure points we 
get a triangle $\Delta_2$, mixtures of four pure points form a tetrahedron 
$\Delta_3$, etc. 

The individuality of a convex set is expressed on its boundary. Each point 
on the boundary belongs to a {\sl face}, which is in itself a convex subset. 
To qualify as a face this convex subset must also be such that for all possible 
ways of decomposing any of its points into pure states, these pure states 
themselves belong to the subset. We will see that the boundary of 
${\cal Q}_N$ is quite different from the boundary of the set of 
classical states.

\subsection{Classical case: the probability simplex}

The simplest convex body one can think of is a {\sl simplex} $\Delta_{N-1}$ 
with $N$ pure states at its corners. The set of all classical states forms 
such a simplex, with the probabilities $p_i$ telling us how much of the $i$th 
pure state that has been mixed in. The simplex is the only convex set which 
is such that a given point can be written as a mixture of pure states in 
one and only one way.   
 
The number $r$ of non--zero components of the vector $\vec{p}$ 
is called the rank of the state. A state of rank one
is pure and corresponds to a corner of the simplex.
Any point inside the simplex $\Delta_{N-1}$ has full rank, $r=N$. 
The boundary of the set of classical states is formed by states
with rank smaller than $N$. Each face is itself a simplex 
$\Delta_{r-1}$. Corners and edges are special cases of faces. 
A face of dimension one less than that of the set itself is 
called a {\sl facet}. 

It is natural to think of the simplex as a regular simplex, with all 
its edges having length one. This can always be achieved, by defining 
the distance between two probability vectors $\vec{p}$ and $\vec{q}$ as 

\begin{equation} D[\vec{p},\vec{q}] = \sqrt{\frac{1}{2}\sum_{i=1}^N(p_i-q_i)^2} \ . 
\end{equation}

\noindent The geometry is that of Euclid. With this geometry in place 
we can ask for the {\sl outsphere}, the smallest sphere that 
surrounds the simplex, and the {\sl insphere}, the largest sphere inscribed 
in it. Let the radius of the outsphere be $R_N$ and that of the insphere 
be $r_N$. One finds that $R_N/r_N = N-1$.  

\subsection{The Bloch ball}

Another simple example of a convex set is a three dimensional ball. 
The pure states sit on its surface, and each such point is a zero 
dimensional face. There are no higher dimensional faces (unless we 
count the entire ball as a face). Given a point that is not pure it 
is now possible to decompose it in infinitely many ways as a mixture 
of pure states.

Remarkably this ball is the space of states ${\cal Q}_2$ of a single 
{\sl qubit}, the simplest quantum mechanical state space. 
For concreteness introduce the Pauli matrices 
$\sigma_1={0\, 1 \choose 1 \, 0}$, 
$\sigma_2={0\,  -{\rm i} \choose {\rm i} \, \, \, 0}$, 
$\sigma_3={1\, \, \, 0 \choose 0 \, -1}$.
These three matrices form an orthonormal basis for the set of traceless 
Hermitian matrices of size two, or in other words for the Lie algebra 
of $SU(2)$.
If we add the identity matrix 
$\sigma_0={\mathbbm 1} = {1\, 0 \choose 0 \, 1}$, 
we can expand an arbitrary state $\rho$ in this basis as 
\begin{equation} 
\label{Bloch2}
\rho = \frac{1}{2}{\mathbbm 1} + \sum_{i=1}^{3}  {\tau}_i{\sigma}_i \ ,
\end{equation}
where the expansion coefficients are 
$\tau_i = {\rm Tr} \rho \sigma_i/2$. 
These three numbers are real since the matrix $\rho$ is Hermitian.
The  three dimensional vector  ${\vec \tau}=(\tau_1,\tau_2,\tau_3)$
is called the {\sl Bloch vector} (or coherence vector). If $\vec{\tau} = 0$ 
we have the {\sl maximally mixed state}. Pure states are represented 
by projectors, $\rho = \rho^2$.  

\begin{figure}[htbp]
\begin{center}
\includegraphics[width=0.25\textwidth,viewport=0 0 320 413,clip=]%
{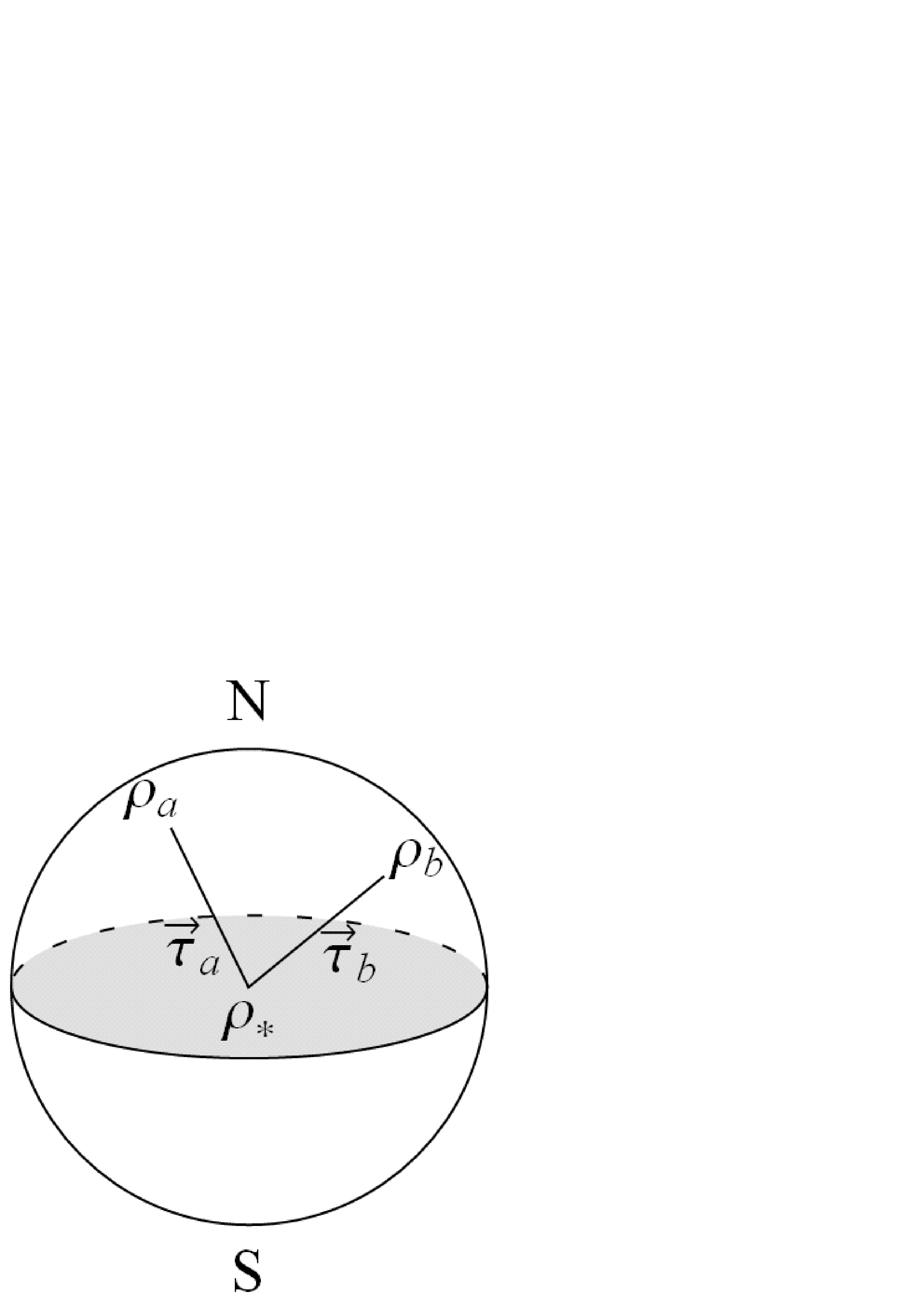}
\caption{The set of mixed states of a qubit forms 
the {\sl Bloch ball} with pure states at the boundary 
and the maximally mixed state $\rho_*=\frac{1}{2}{\mathbbm 1}$
at its center:
The Hilbert--Schmidt distance between any two
states is the length of the difference between their Bloch vectors, 
$||\vec{\tau}_a-\vec{\tau}_b||$.}
\label{fig:Blochball}
\end{center}
\end{figure}

Since the Pauli matrices are traceless the coefficient $\frac{1}{2}$
standing in front of the identity matrix assures that Tr$\rho=1$, but 
we must also ensure that all eigenvalues are non-negative. By computing 
the determinant we find that this is so if and only if the length of 
the Bloch vector is bounded, $||\vec\tau||^2 \le 1$. Hence ${\cal Q}_2$ is 
indeed a solid ball, with the pure states forming its surface---the 
{\sl Bloch sphere}.

A simple but important point is that the set of classical 
states $\Delta_1$, which is just a line segment in this case, 
sits inside the Bloch ball as one of its diameters. This goes 
for any diameter, since we are free to regard any two commuting 
projectors as our classical bit. Two commuting projectors sit at 
antipodal points on the Bloch sphere. To ensure that the distance 
between any pair of antipodal equals one we define the distance 
between two density matrices $\rho_A$ and $\rho_B$ to be 

\begin{equation} 
D_{\rm HS}(\rho_A,\rho_B) \ = \ 
\sqrt{ \frac{1}{2}\mbox{Tr}[(\rho_A-\rho_B)^2] }  \ . 
\label{HSdist}
\end{equation}
This is known as the {\sl Hilbert-Schmidt distance}. Let us 
express this in the Cartesian coordinate system provided 
by the Bloch vector,

\begin{equation} 
\label{HStau}
D_{\rm HS}[\rho_A,\rho_B] \ = \ 
\sqrt{ \sum_{i=1}^3  (\tau^A_i -\tau^B_i)^2 } \ = \
||\vec{\tau}^A - \vec{\tau}^B||\ .
\end{equation}
This is the Euclidean notion of distance.

\subsection{Quantum case: ${\cal Q}_N$}

When $N > 2$ the quantum state space is no longer a solid ball. It 
is always a convex set however. Given two density matrices, 
that is to say two positive hermitian matrices 
$\rho,\sigma \in {\cal Q}_N$.
It is then easy to see that any convex combination of these two states, 
$a \rho+ (1-a)\sigma \in {\cal Q}_N$ where $a\in [0,1]$, must be a 
positive matrix too, and hence it belongs to ${\cal Q}_N$. 
This shows that the set of quantum states is convex. For all $N$ the 
face structure of the boundary can be discussed in a unified way. 
Moreover it 
remains true that ${\cal Q}_N$ is swept out by rotating a classical 
probability simplex $\Delta_{N-1}$ in $\mathbbm R^{N^2-1}$, but for 
$N > 2$ there are restrictions on the allowed rotations.  

To make these properties explicit we start with the observation that 
any density matrix can be represented as a convex combination
of pure states
\begin{equation} 
\label{convcomb}
\rho = \sum_{i=1}^{k} p_i \, |\phi_i \rangle \langle \phi_i|,
\end{equation}
where $\vec{p}= (p_1,p_2,\dots, p_k)$ is a probability vector.
In contrast to the classical case there exist infinitely many 
decompositions of any mixed state $\rho \ne \rho^2$. 
The number $k$ can be arbitrarily large,
and many different choices can be made for the pure states 
$|\phi_i \rangle$. But there does exist a distinguished decomposition.
Diagonalising the density matrix we 
find its eigenvalues $\lambda_i\geq 0$ and eigenvectors $|\psi_i\rangle$.
This allows us to write the eigendecomposition of a state, 
\begin{equation} \rho = 
\sum_{j=1}^{N} \lambda_j |\psi_j \rangle \langle \psi_j| \ .
\end{equation}
The number $r$ of non-zero components of the probability vector $\vec{\lambda}$ 
is called the rank of the state $\rho$, 
and does not exceed $N$. This is the usual definition of the rank 
of a matrix, and by happy accident it agrees with the definition of 
rank in convex set theory: the {\sl rank} of a point in a convex set is 
the smallest number of pure points needed to form the given point 
as a mixture. 

Consider now a general convex set in $d$ dimensions.
Any point belonging to it 
can be represented by a convex combination of not more than $d+1$
extremal states.
Interestingly, ${\cal Q}_N$  has a peculiar geometric structure
since any given density operator $\rho$ can be represented by 
a combination of not more than $N$ pure states, 
which is much smaller than $d+1=N^2$. In Hilbert space these $N$ pure 
states are the orthogonal eigenvectors of $\rho$. 
If we adopt the Hilbert-Schmidt 
definition of distance (\ref{HSdist})  
they form a copy of the 
classical state space, the regular simplex $\Delta_{N-1}$. 

Conversely, 
every density matrix can be reached from a diagonal density matrix 
by means of an $SU(N)$ transformation. Such transformations form 
a subgroup of the rotation group 
$SO(N^2-1)$. Therefore any density matrix can be obtained by rotating 
a classical probability simplex around the maximally mixed state, which 
is left invariant by rotations. However, when $N > 2$ $SU(N)$ is a 
proper subgroup of $SO(N^2-1)$, which is why ${\cal Q}_N$ forms a solid 
ball only if $N = 2$. The relative sizes of the outsphere and 
the insphere are still related by $R_N/r_N = N-1$. 

\begin{figure}
\begin{center}
\includegraphics[width=0.44\textwidth,viewport=14 14 529 439,clip=]%
{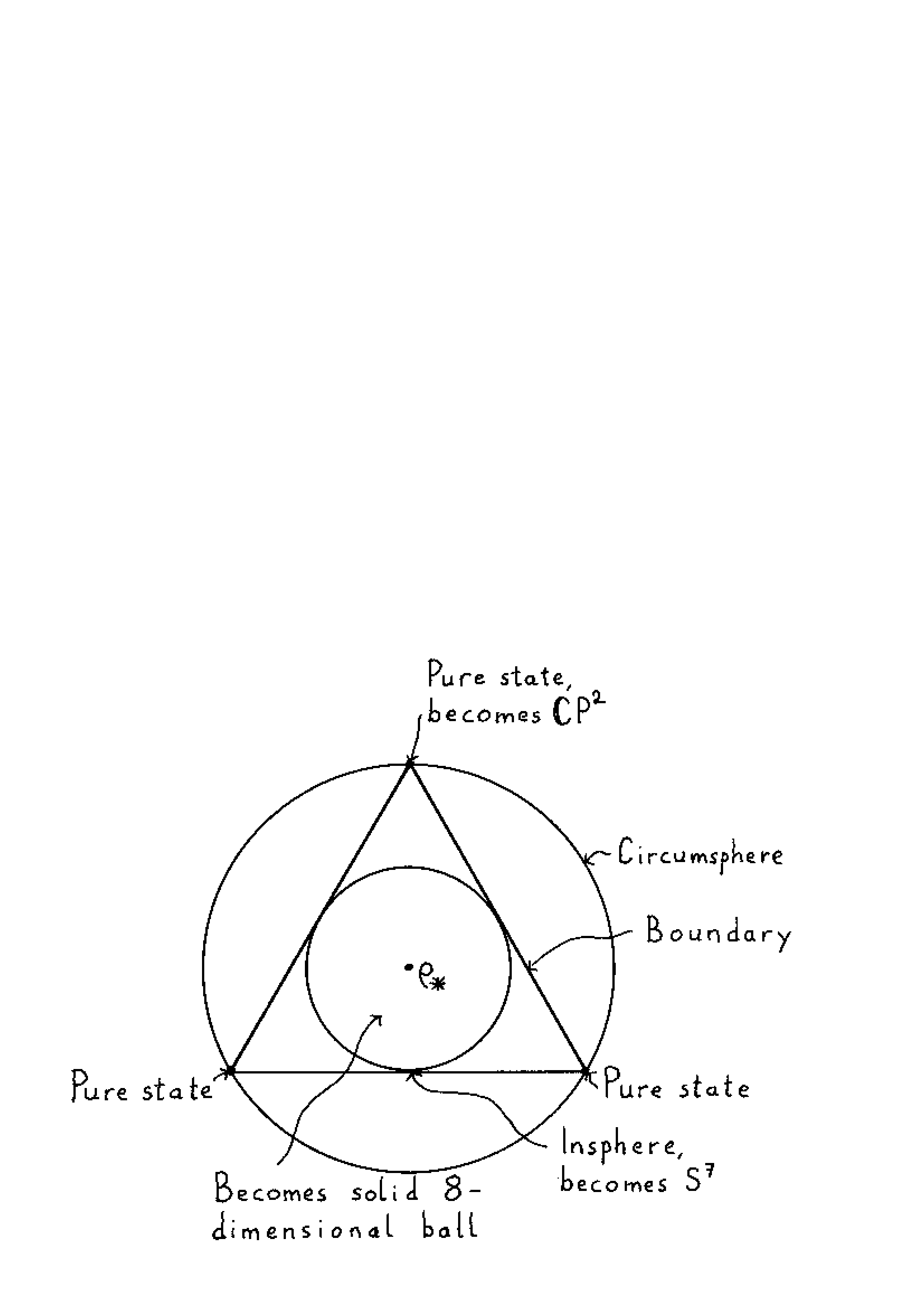}
\end{center}
\label{fig:densthree}
\caption{
 The set ${\cal Q}_3$ of quantum states of a qutrit 
contains positive semi-definite matrices with spectrum
from the simplex $\Delta_2$ of classical states. 
The corners of the triangle become the $4D$ set of pure states, 
the edges lead to the  $7D$ boundary  $\partial{\cal Q}_3$,
while interior of the triangle gives
the interior of the $8D$ convex body.
The set ${\cal Q}_3$ is inscribed inside a $7$--sphere of radius
$R_3=\sqrt{2/3}$ and it contains an $8$--ball of radius $r_3=1/\sqrt{6}$.}
\end{figure}

The boundary of the set ${\cal Q}_N$ shows some similarities with that of its 
classical cousin. It consists of all matrices whose rank is smaller 
than $N$. There will be faces of rank 1 (the pure states), of rank 2 
(in themselves they are copies of ${\cal Q}_2$), and so on up to 
faces of rank $N-1$ (copies of ${\cal Q}_{N-1}$). Note that there are 
no hard edges: the minimal non-extremal faces are solid three dimensional 
balls. The largest faces have a dimension much smaller than the 
dimension of the boundary of ${\cal Q}_N$. As in the classical 
case, any face can be described as the intersection of the convex set 
with a bounding hyperplane in the container space. In technical language 
one says that all faces are exposed. Note also that every 
point on the boundary belongs to a face that is tangent to the insphere. 
This has the interesting consequence that the area $A$ of the 
boundary is related to the volume $V$ of the body by 
\begin{equation} \frac{rA}{V} = d \ , 
\label{eq:rAdV}
\end{equation}
where $r$ is the radius of the insphere and $d$ is the dimension 
of the body (in this case $d = N^2-1$) \cite{SBZ06}. Incidentally 
the volume of ${\cal Q}_N$ is known explicitly \cite{ZS03}. 

There are differences 
too. A typical state on the boundary has rank $N-1$, and any two such 
states can be connected with a curve of states such that all states 
on the curve have the same rank. In this sense ${\cal Q}_N$ is more 
like an egg than a polytope \cite{Marmo}. 

We can regard the set of $N$ by $N$ matrices as a vector space
(called Hilbert-Schmidt space),
endowed with the scalar product 
\begin{equation} \langle A | B \rangle_{\rm HS} = 
\frac{1}{2}{\rm Tr} A^{\dagger} B \ . 
\label{HSproduct} \end{equation} 
The set of hermitian matrices with unit trace is not a vector 
space as it stands, but it can be made into one by separating out the 
traceless part. Thus we can represent a density matrix as  

\begin{equation} \rho = \frac{1}{N}{\mathbbm 1} + u \ , \label{vectors} 
\end{equation}
\noindent where $u$ is traceless. The set of traceless matrices is an 
Euclidean subspace of Hilbert-Schmidt space, and the Hilbert-Schmidt distance (\ref{HSdist}) arises 
from this scalar product. In close analogy to eq. (\ref{Bloch2}) we can 
introduce a basis for the set of traceless matrices, and write the 
density matrix in the  
{\sl generalized Bloch vector} representation,
\begin{equation} 
\label{bloch3}
\rho = \frac{1}{N}{\mathbbm 1} + \sum_{i=1}^{N^2-1} u_i{\gamma}_i 
\ .
\end{equation}
Here $\gamma_i$ are hermitian basis vectors. The components $u_i$ must be 
chosen such that $\rho$ is a positive definite matrix.  

\subsection{Dual and self-dual convex sets}

Both the classical and the quantum state spaces have the remarkable 
property that they are {\sl self-dual}. But the word duality has many 
meanings. 
In projective geometry the dual of a point is a plane.  
If the point is represented by a vector ${\vec x}$, we can 
define the dual plane as the set of vectors ${\vec y}$ such that 

\begin{equation}
 {\vec x}\cdot {\vec y} = - 1 \ . 
\label{eq:dualplane}
\end{equation}
The dual of a line is the intersection of a one-parameter family of 
planes dual to the points on the line. This is in itself a line. The dual 
of a plane is a point, while the dual of a curved surface is another 
curved surface---the envelope of the planes that are dual to the points 
on the original surface. To define the dual of a convex body with a 
given boundary we change the definition slightly, and include all  
points on one side of the dual planes in the dual. Thus the {\sl dual} 
$X^*$ of a convex body $X$ is defined to be 

\begin{equation}
X^*=\{{\vec x}\mid 
1 + {\vec x}\cdot {\vec y} \; \geq \; 0\ \forall {\vec y}\in X\} \ . 
\label{eq:dualgen}
\end{equation}
The dual of a convex body including the origin is the intersection of 
half-spaces $\{{\vec x}\mid 1 + {\vec x}\cdot {\vec y} \geq 0\}$ for
extremal points ${\vec y}$ of $X$ \cite{R70}. If we enlarge a convex body
the conditions on the dual become more stringent, and hence the dual
shrinks. The dual of a sphere centred at the origin is again a sphere,
so a sphere (of suitable radius) is self-dual. The dual of a cube is an 
octahedron. The dual of a regular tetrahedron is another copy of the
original tetrahedron, possibly of a different size. Hence this is a
self-dual body. Convex subsets $F\subset X$ are mapped to subsets of
$X^*$ by
\begin{equation}
F\;\mapsto\;\widehat{F}:=
\{{\vec x}\in X^*\mid 1 + {\vec x}\cdot {\vec y} = 0\,\forall {\vec y}\in F\} \ .
\label{eq:dual_face}
\end{equation}
Geometrically, $\widehat{F}$ equals $X^*$ intersected with the dual affine
space (\ref{eq:dualplane}) of the affine span of $F$. 
If the origin lies in the interior of the convex body $X$ then
$F\mapsto\widehat{F}$ is a one-to-one inclusion-reversing correspondence
between the exposed faces of $X$ and of $X^*$ \cite{Gr03}.
If $X$ is a tetrahedron, then vertices and faces are exchanged, while
edges go to edges.

%

What we need in order to prove the self-duality of ${\cal Q}_N$ is 
the key fact that a hermitian and unit trace matrix $\sigma$ is a 
density matrix if and only if 

\begin{equation} {\rm Tr} \sigma \rho \ \geq \  0 \label{rhosigma} 
\end{equation}
for all density matrices $\rho$. It will be convenient to think 
of a density matrix $\rho$ as represented by a ``vector'' $u$, 
as in eq. (\ref{vectors}). As a direct consequence of eq. 
(\ref{rhosigma}) the set of quantum states ${\cal Q}_N$ is 
self-dual in the precise sense that 
 
\begin{equation}
{\cal Q}_N-\id/N=\{u\mid 1/N +{\rm Tr}(u v)\geq 0\,\forall v\in {\cal Q}_N-\id/N\} .
\label{eq:dualmN}
\end{equation}
In this equation the trace is to be interpreted as a scalar product in 
a vector space. Duality (\ref{eq:dual_face}) exchanges faces of rank $r$
(copies of ${\cal Q}_r$) and faces of rank $N-r$ (copies of ${\cal Q}_{N-r}$).

Self-duality is a key property of state spaces \cite{Wilce, Ududec}, 
and we will use it extensively when we discuss projections and
cross-sections of ${\cal Q}_N$. 
This notion is often introduced in the larger vector space 
consisting of all hermitian matrices, with the origin at the zero 
matrix. The set of positive semi-definite matrices forms a cone 
in this space, with its apex at the origin. It is a cone because 
any positive semi-definite matrix remains positive semi-definite 
if multiplied by a positive real number. This defines the rays of 
the cone, and each ray intersects the set of unit trace matrices 
exactly once. The dual of this cone is the set of all matrices 
$a$ such that Tr$ab \geq 0$ for all matrices $b$ within the 
cone---and indeed the dual cone is equal to the original, so 
it is self-dual. 

\section{An apophatic approach to the qutrit}
\label{sec3}

For $N = 3$ we are dealing with the states of the {\sl qutrit}.  
The Gell-Mann matrices are a standard choice \cite{Goyal} for the 
eight matrices $\gamma_i$, and the expansion 
coefficients are $\tau_i = \frac{1}{2} {\rm Tr} \rho \gamma_i$. Unfortunately, although 
the sufficient conditions for $\vec{\tau}$ to represent 
a state are known \cite{AMM97,Ki03,KK05}, they do 
not improve much our understanding of the geometry of ${\cal Q}_3$.

We know that the set of pure states has 4 real dimensions, 
and that the faces of ${\cal Q}_3$ are copies of the 3D Bloch ball, 
filling out the 7 dimensional boundary. The centres of these balls 
touch the largest inscribed sphere of ${\cal Q}_3$. But what does it 
all really look like?

We try to answer this question by presenting
some 3D objects, and explaining why they cannot serve as models
of ${\cal Q}_3$. Apart from the fact that our objects are not eight 
dimensional,
all of them lack some other features of the set of quantum states.

\begin{figure}[htbp]
\begin{center}
\includegraphics[width=0.31\textwidth,viewport=13 13 1590 1726,clip=]%
{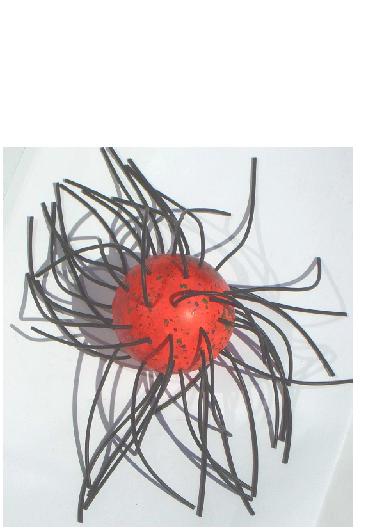}
\caption{Apophatic approach: this object is
{\sl not} a good model of the set ${\cal Q}_3$ 
as it is not a convex set.}
\label{fig1a}
\end{center}
\end{figure}

\begin{figure}[htbp]
\begin{center}
\includegraphics[width=0.31\textwidth,viewport=13 13 1490 1504,clip=]%
{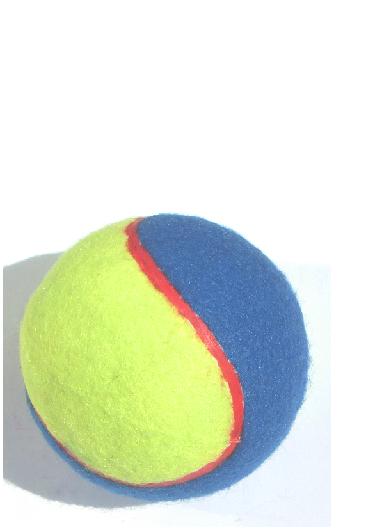}
\end{center}
\caption{The set ${\cal Q}_3$ 
is {\sl not} a ball...}
\label{fig2}
\end{figure}

\begin{figure}[htbp]
\begin{center}
\includegraphics[width=0.31\textwidth,viewport=13 13 1651 1605,clip=]%
{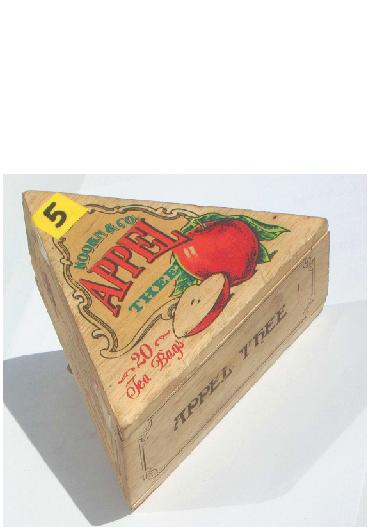}
\end{center}
\caption{The set ${\cal Q}_3$ 
is {\sl not} a polytope...}
\label{fig3}
\end{figure}

Fig.~\ref{fig1a} presents a hairy set which is nice but not convex.
Fig.~\ref{fig2} shows a ball, and we know that 
${\cal Q}_3$ is not a ball.
It is not a polytope either,
so the polytope shown in Fig.~\ref{fig3} 
cannot model the set of quantum states.

\begin{figure}[htbp]
\begin{center}
\includegraphics[width=0.31\textwidth,viewport=13 65 1086 1245,clip=]%
{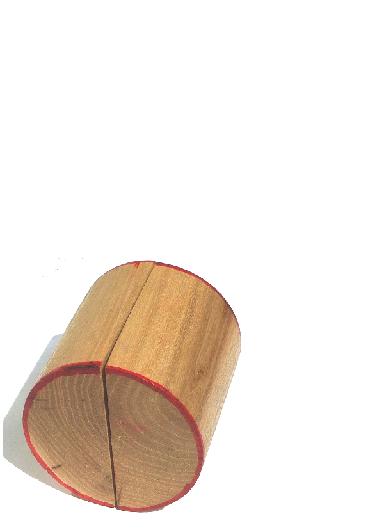}
\end{center}
\caption{The set
of pure states in ${\cal Q}_3$ is connected,
but for the cylinder the pure states form two circles.}
\label{fig4}
\end{figure}

\begin{figure}[htbp]
\begin{center}
\includegraphics[width=0.31\textwidth,viewport=19 13 1401 1191,clip=]%
{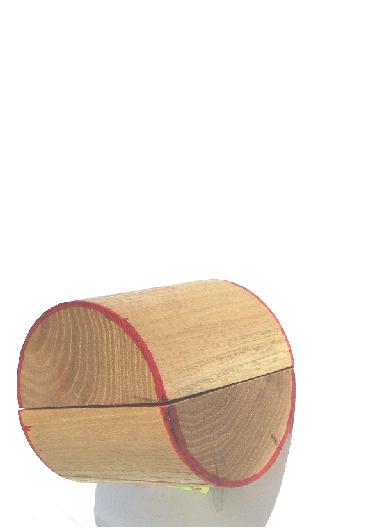}
\end{center}
\caption{
This is now the convex hull of a single space curve, but 
one cannot inscribe copies of the classical set $\Delta_2$ in it.}
\label{fig5}
\end{figure}

Let us have a look at the cylinder shown in Fig.~\ref{fig4},
and locate the extremal points of the convex body shown. This 
subset consists of the two circles 
surrounding both bases. This is a disconnected set, 
in contrast to the connected set of pure quantum states.
However, if one splits the cylinder into two halves
and rotates one half by $\pi/2$ as shown 
in Fig.~\ref{fig5}, one obtains a body with a connected 
set of pure states.
A similar model can be obtained by taking the
convex hull of the seam of a tennis ball:
the one dimensional seam  contains the extremal points of this
set and forms a connected set.

Thus the seam of the tennis ball (look again at Fig. \ref{fig2})
corresponds to the $4D$ connected set of pure states of $N=3$ quantum system.
The convex hull of the seam forms a $3D$ object which is easy to visualize, 
and serves as our first rough model of the solid $8D$ body ${\cal Q}_3$
of qutrit states.
However, a characteristic feature of the latter is that each one of 
its points belongs to a cross-section which is an equilateral triangle 
$\Delta_2$. (This is the eigenvector decomposition.) 
The convex set determined by the seam of the tennis ball, 
and the set shown in Fig. \ref{fig5}, do not have this property.

\begin{figure}
\begin{center}
\includegraphics[width=0.34\textwidth,viewport=14 14 311 266,clip=]%
{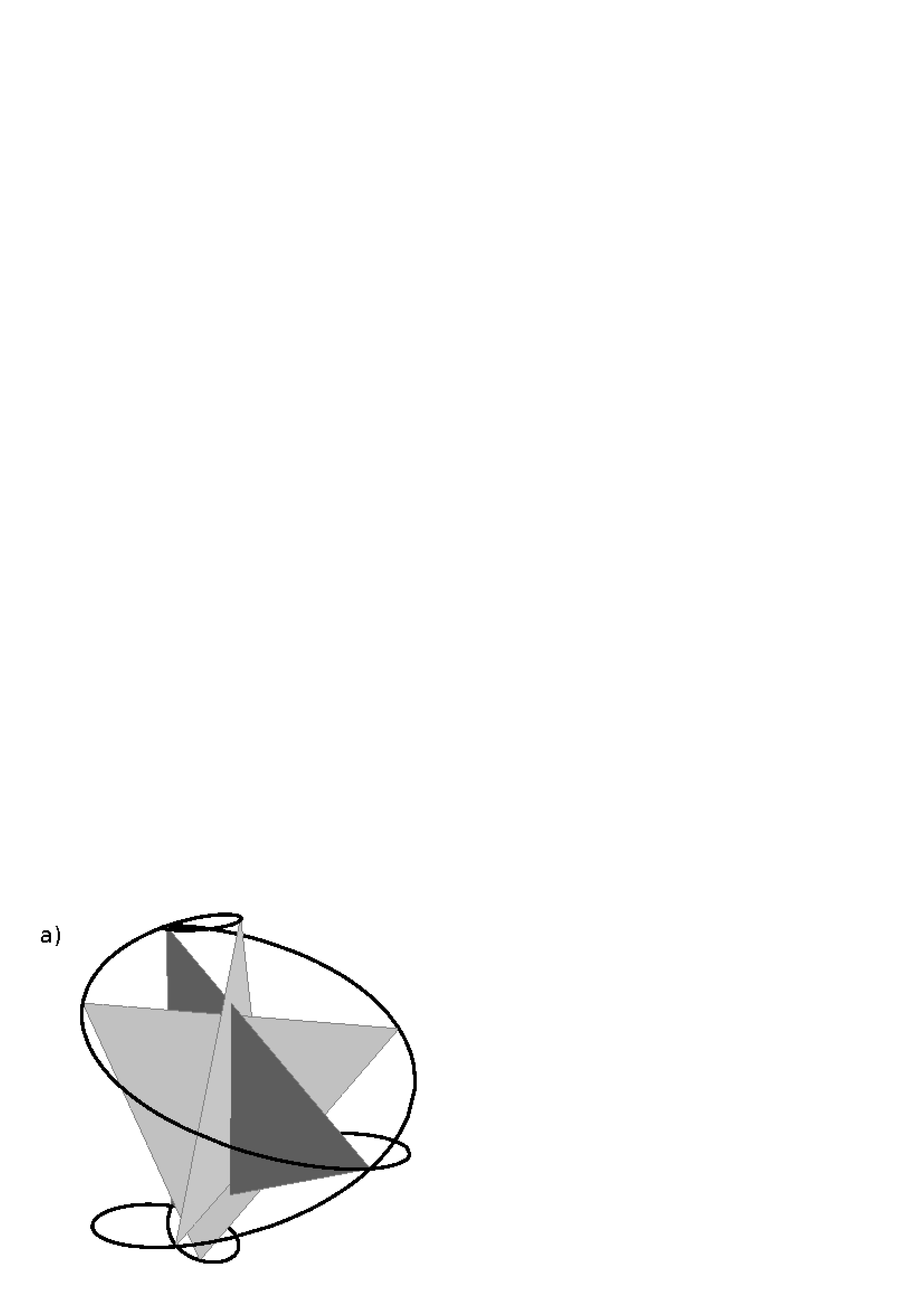}
\includegraphics[width=0.34\textwidth,viewport=14 14 324 295,clip=]%
{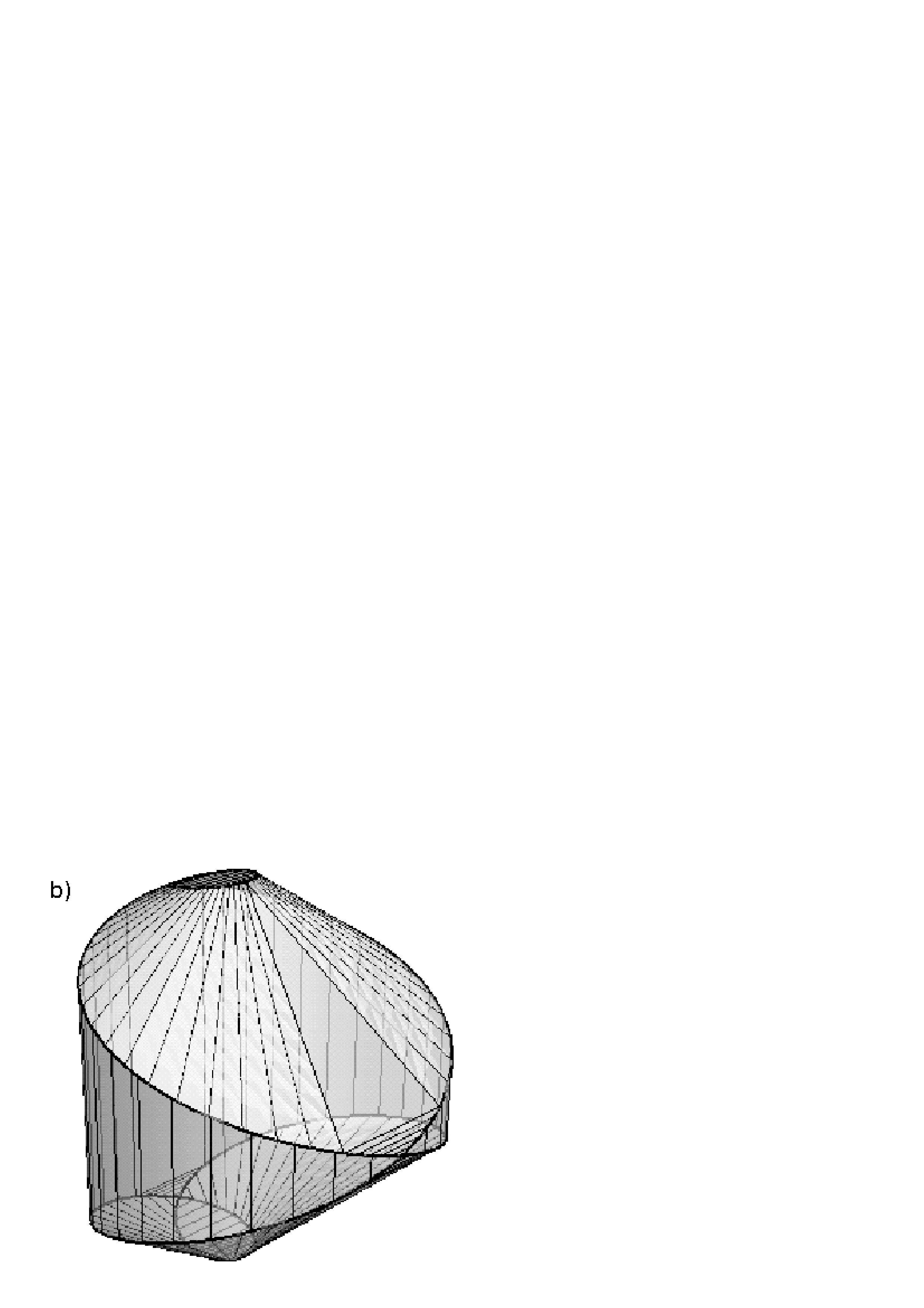}
\end{center}
\caption{\label{fig:hull_curve}
a) The space curve ${\vec x}(t)$ modelling pure quantum states
is obtained by rotating an equilateral triangle 
according to Eq. (\ref{curvec})
  ---three positions of the triangle are shown);
 b) The convex hull $C$ of the curve models the set of all quantum states.}
\end{figure}

As we have seen ${\cal Q}_3$ can be obtained if we take an equilateral
triangle $\Delta_2$ and subject it to $SU(3)$ rotations in eight
dimensions. We can try to do something similar in three dimensions. 
If we rotate a triangle along one of its bisections we obtain a cone, 
for which the set of extremal states consists of a circle and an apex 
(see Fig. \ref{fig:steiner_cayley} b)), a disconnected set.
We obtain a better model  
if we consider the space curve
\begin{equation}
{\vec x}(t) \; = \;
 \bigl(\cos(t)\cos(3t),\ \cos(t)\sin(3t),\ -\sin(t) \bigr)^{\rm T} \; .
\label{curvec}
\end{equation}
Note that the curve is closed, ${\vec x}(t)={\vec x}(t+2\pi)$,
and belongs to the unit sphere, $||{\vec x}(t)||=1$.
Moreover 
\begin{equation} ||{\vec x}(t)-{\vec x}(t+\tfrac{1}{3}2\pi)||=\sqrt{3} \end{equation} 
for every value of $t$.
Hence every point ${\vec x}(t)$ belongs to an equilateral triangle with 
vertices at 
\[
{\vec x}(t),\qquad {\vec x}(t+\tfrac{1}{3}2\pi), \qquad \text{and} \qquad
{\vec x}(t+\tfrac{2}{3}2\pi) \ . 
\]
They span a plane including the $z$-axis 
for all times $t$.
During the time  $\Delta t=\tfrac{2\pi}{3}$ this plane makes a full 
turn about the $z$-axis, while the triangle rotates by the angle 
$2\pi/3$ within the plane---so the triangle has returned to a congruent 
position. The curve ${\vec x}(t)$ is shown in 
Fig.~\ref{fig:hull_curve} a) together with 
exemplary positions of the rotating triangle, and Fig.~\ref{fig:hull_curve} b) 
shows its convex hull $C$. This convex hull is symmetric 
under reflections in the $(x$-$y)$ and $(x$-$z)$ planes. 
Since the set of pure states is connected this is our best model 
so far of the set of quantum pure states, although the likeness is not 
perfect. 

It is interesting to think a bit more about the boundary of $C$. There are 
three flat faces, two triangular ones and one rectangular. 
The remaining part of the boundary consists of ruled surfaces: they are 
curved, but contain one dimensional faces (straight lines). The boundary 
of the set shown in Fig. \ref{fig5} has similar properties. 
The ruled surfaces of $C$ have an analogue in the boundary of the set of
quantum states ${\cal Q}_3$, we have already noted that a generic point
in the boundary of ${\cal Q}_3$ belongs to a copy of ${\cal Q}_2$
(the Bloch ball), arising as the intersection of ${\cal Q}_3$ with a
hyperplane. The flat pieces of $C$ have no analogues in the boundary of
${\cal Q}_3$, apart from Bloch balls (rank two) and pure states (rank one)
no other faces exist. 
  
Still this model is not perfect: Its set of pure states has
self-intersections. Although it is created by rotating a triangle, the
triangles are not cross-sections of $C$. It 
is not true that every point on the boundary belongs to a face 
that touches the largest inscribed sphere,
as it happens for the set of quantum states \cite{SBZ06}. Indeed its 
boundary is not quite what we want it to be, in particular it has 
non-exposed faces---a point to which we will return. Above all this is 
not a self-dual body. 

\section{A constructive approach}

The properties of the eight-dimensional convex set ${\cal Q}_3$ might
conflict if we try to realize them in dimension three. Instead of looking
for an ideal three dimensional model we shall thus use a complementary
approach. To reduce the dimensionality of the problem we investigate 
cross-sections of the $8D$ set ${\cal Q}_3$ with a plane of dimension
two or three, as well as its orthogonal projections on these planes---the 
shadows cast by the body on the planes, when illuminated by a very distant
light source. Clearly the cross-sections will always be contained in the 
projections, but in exceptional cases they may coincide. 

What kind of cross-sections arise? In the classical case it is 
known that every convex polytope arises as a cross-section of a 
simplex $\Delta_{N-1}$ of sufficiently high dimension \cite{Gr03}. 
It is also 
true that every convex polytope arises as the projection of a 
simplex. But what are the cross-sections 
and the projections of ${\cal Q}_N$? There has been considerable 
progress on this question recently. The convex set is said to be a 
{\sl spectrahedron} if it 
is a cross-section of a cone of semi-positive definite 
matrices of some given size. In the branch of mathematics known as 
convex algebraic geometry one asks what kind of convex bodies that 
can be obtained as projections of spectrahedra. Surprisingly, the convex 
hull of any trigonometric space curve in three dimensions can be 
so obtained \cite{Henrion}. 
This includes our set $C$, which 
can be shown to be a projection of an $8$-dimensional cross-section 
of the $35D$ set ${\cal Q}_6$ of quantum states of size $N=6$. 
We do so in Appendix~\ref{app:C}.

\subsection{The duality between projections and cross-sections}

\begin{figure}
\begin{center}
\includegraphics[width=0.34\textwidth,viewport=0 0 240 240,clip=]%
{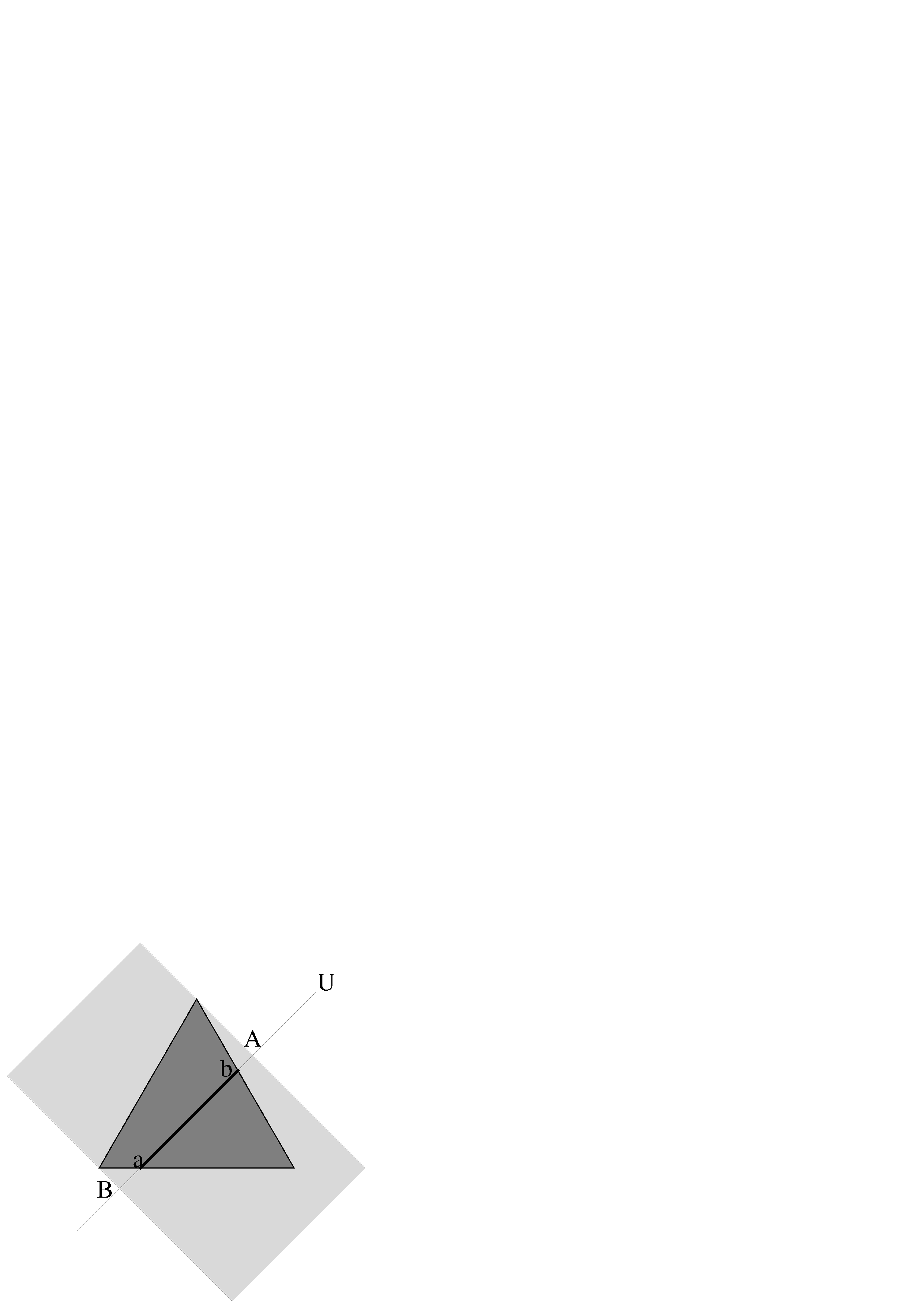}
\end{center}
\caption{\label{fig:projection}The triangle is self-dual. We intersect
it with a one-dimensional 
subspace through the centre, $U$, and obtain a cross-section extending 
from $a$ to $b$. The dual 
of this line in the plane is a 2-dimensional strip, and when we project this onto $U$ 
we obtain a projection extending from $A$ to $B$, which is dual to the cross-section 
within $U$.}
\end{figure}

In the vector space of traceless hermitian matrices we choose a linear 
subspace $U$. The intersection of the convex body ${\cal Q}_N$ of
quantum states with the subspace $U + \id /N$ through the maximally
mixed state $\id /N$ is the cross-section $S_U$, and the orthogonal
projection of ${\cal Q}_N$ down to $U$ is the projection $P_U$. There
exists a beautiful relation between 
projections and cross-sections, holding for self-dual convex bodies 
such as the classical 
and the quantum state spaces \cite{Weis_cs}. For them  
cross-sections and projections 
are dual to each other, in the sense that 

\begin{equation}
S_U-\id/N=\{u\mid 1/N +{\rm Tr}(u v)\geq 0\ \forall v\in P_U\} \ \ 
\label{eq:cross}
\end{equation}
and 

\begin{equation}
P_U=\{u\mid 1/N +{\rm Tr}(u v)\geq 0\ \forall v\in S_U-\id/N\} \ . 
\label{eq:proj}
\end{equation}
This is best explained in a picture (namely Fig. \ref{fig:projection}). 
A special case of these dualities is the self-duality of the full 
state-space, eq. (\ref{eq:dualmN}). 

\begin{figure}
\begin{center}
\includegraphics[width=0.20\textwidth,viewport=0 0 179 180,clip=]%
{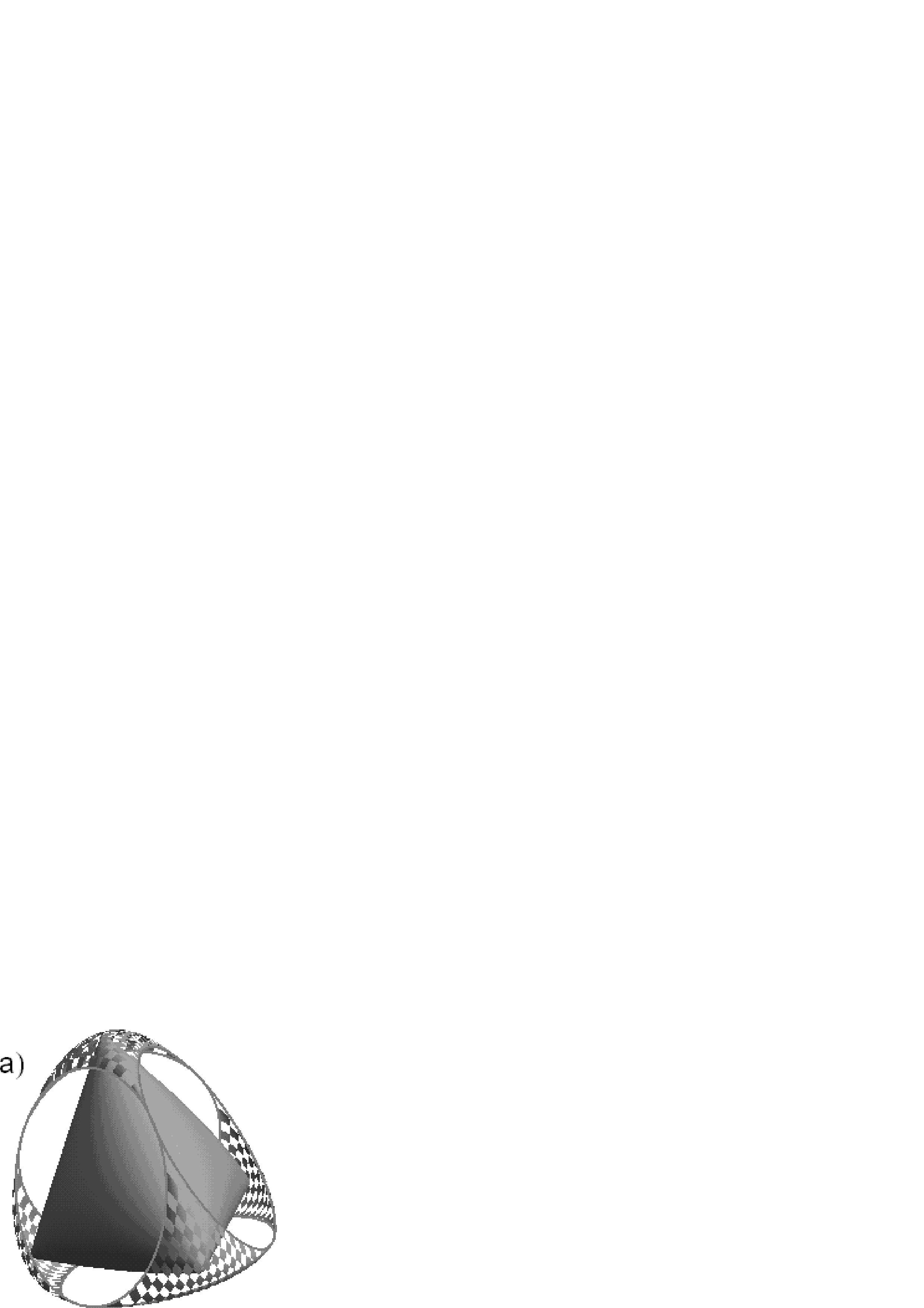}%
\hspace{0.5cm}
\includegraphics[width=0.20\textwidth,viewport=0 0 190 185,clip=]%
{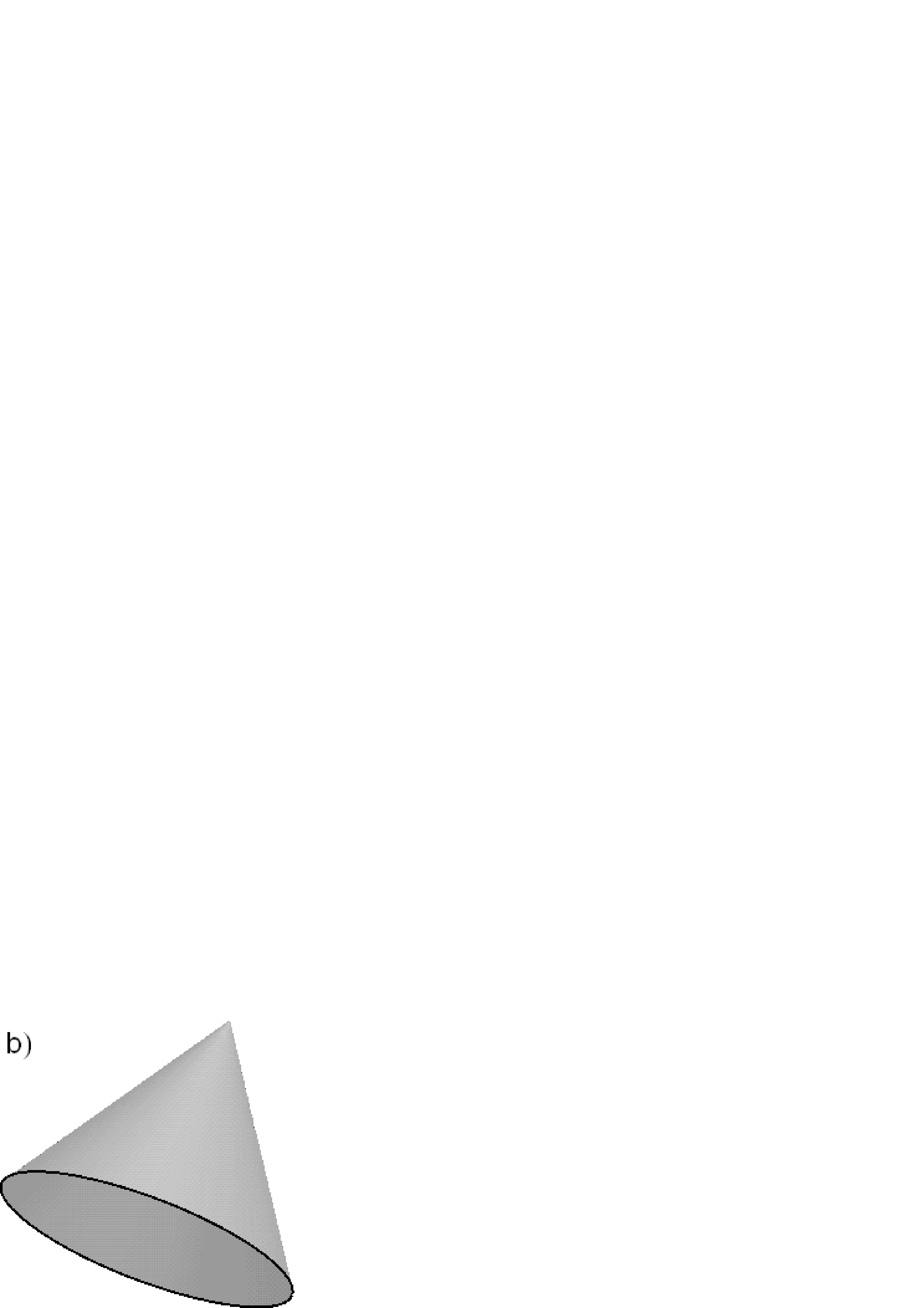}
\end{center}
\caption{\label{fig:steiner_cayley}
a) The cross-section $S_U-\id/3$ defined in (\ref{eq:cayley}) of the
qutrit quantum states ${\cal Q}_3$ is drawn inside the projection
$P_U$ of ${\cal Q}_3$. b) The cone is self-dual, it is a cross-section
and a projection of ${\cal Q}_3$ with $S_U-\id/3=P_U$.}
\end{figure}

Let us look at two examples for ${\cal Q}_3$, choosing the vector space 
$U$ to be three dimensional. In Fig.\ref{fig:steiner_cayley} a) we show 
the cross-section containing all states of the form 

\begin{equation} \rho = \left( \begin{array}{ccc} 1/3 & x & y \\ 
x & 1/3 & z \\ y & z & 1/3 \end{array} \right) \ , \hspace{5mm} 
\rho \geq 0 \ . 
\label{eq:cayley}\end{equation}
They form an overfilled tetrapak cartoon \cite{Bl76}, also known as an 
elliptope \cite{Sturmfels} and an obese tetrahedron \cite{Goyal}. Like 
the tetrahedron it has six straight edges. Its 
boundary is known as Cayley's cubic surface, and it is smooth everywhere 
except at the four vertices. In the picture it is 
surrounded by its dual projection, 
which is the convex hull of a quartic surface known as Steiner's Roman 
surface. To understand the shape of the dual, start with a pair of 
dual tetrahedra (one of them larger than the other). Then we ``inflate'' 
the small tetrahedron a little, so that its 
facets turn into curved surfaces. It grows larger, so its dual must 
shrink---the vertices of the dual become smooth, while the 
facets of the dual will be contained within the original 
triangles. What we see in Fig.\ref{fig:steiner_cayley} a) is a 
``critical'' case, in which the facets of the dual have shrunk to four 
circular disks that just touch each other in six special points. 

In Fig.\ref{fig:steiner_cayley} b) we see the cross-section containing 
all states (positive matrices) of the form 

\begin{equation} \rho = \left( \begin{array}{ccc} 
1/3+z/\sqrt{3} & x -{\rm i}y & 0 \\ 
x+{\rm i}y & 1/3+z/\sqrt{3} & 0 \\ 
0 & 0 & 1/3 - 2z/\sqrt{3}  \end{array} \right) \ . 
\end{equation}
This cross-section is a self-dual set, meaning that the projection to
this 3-dimensional plane coincides with the cross-section. In itself it
is the state space of a real subalgebra of the qutrit obervables. There
exist also two-dimensional self-dual cross-sections, which are simply
copies of the classical simplex $\Delta_2$---the state 
space of the subalgebra of diagonal matrices. 

\subsection{Two-dimensional projections and cross-sections}

To appreciate what we see in cross-sections and projections we 
will concentrate on 2-dimensional screens. 

We can compute 2D projections using the fact that they are dual to 
a cross-section. But we can also use the notion of 
the {\sl numerical range} $W$ of a given operator $A$,
a subset of the complex plane \cite{HJ2,GR97,GPMSZ10}
\begin{equation}
W(A)=\{ z\in {\mathbbm C}: z= {\rm Tr} \rho A, \ \rho \in {\cal Q}_N \} \ .
\label{range2}
\end{equation}
If the matrix $A$ is hermitian its numerical range reduces to a line segment, 
otherwise it is a convex region of the complex plane. To see the connection 
to projections, observe that changing the trace of $A$ gives rise to a 
translation of the whole set, so we may as well fix the trace to equal 
unity. Then we can write for some $\lambda\in{\mathbb C}$

\begin{equation}
 A = \lambda{\mathbbm 1} + u + {\rm i}v \ , \end{equation}
where $u$ and $v$ are traceless hermitian matrices. It follows that 
the set of all possible numerical ranges $W(A)$ 
of arbitrary matrices $A$ of order $N$
is affinely equivalent to the set of 
orthogonal projections of ${\cal Q}_N$ on a $2$-plane \cite{Dunkl, Henrionrange}.
Thus to understand the structure of projections of ${\cal Q}_N$ 
onto a plane it is sufficient to analyze the geometry
of numerical ranges of any operator of size $N$.
For instance, in the simplest case of a matrix $A$
of order $N=2$, its numerical range forms an elliptical disk,
which may reduce to an interval.
These are just possible (not necessarily orthogonal) 
projections of the Bloch ball
${\cal Q}_2$ onto a plane.

\begin{figure}
\begin{center}
\includegraphics[width=0.18\textwidth,viewport=14 14 362 416,clip=]%
{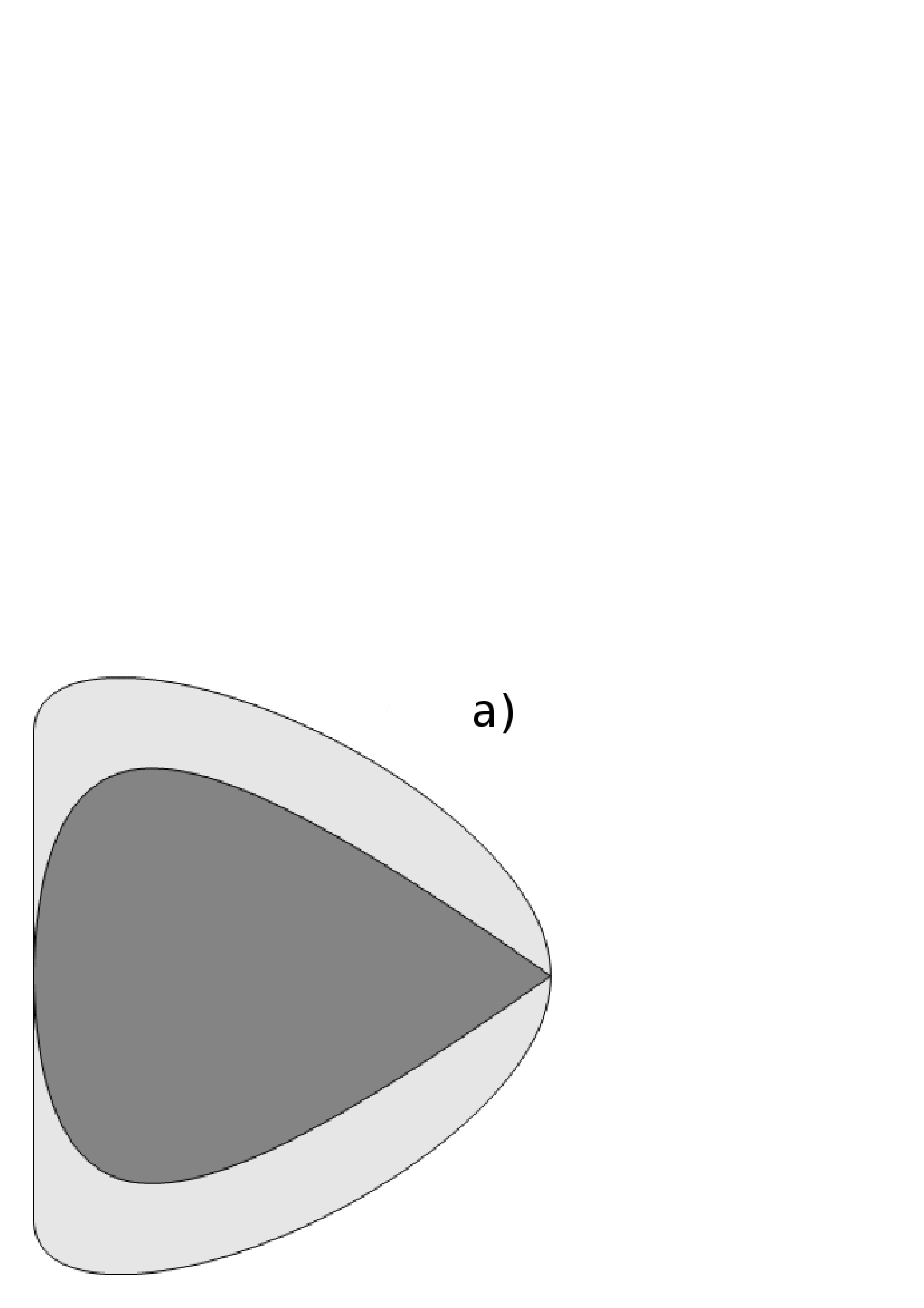}%
\includegraphics[width=0.20\textwidth,viewport=14 14 397 357,clip=]%
{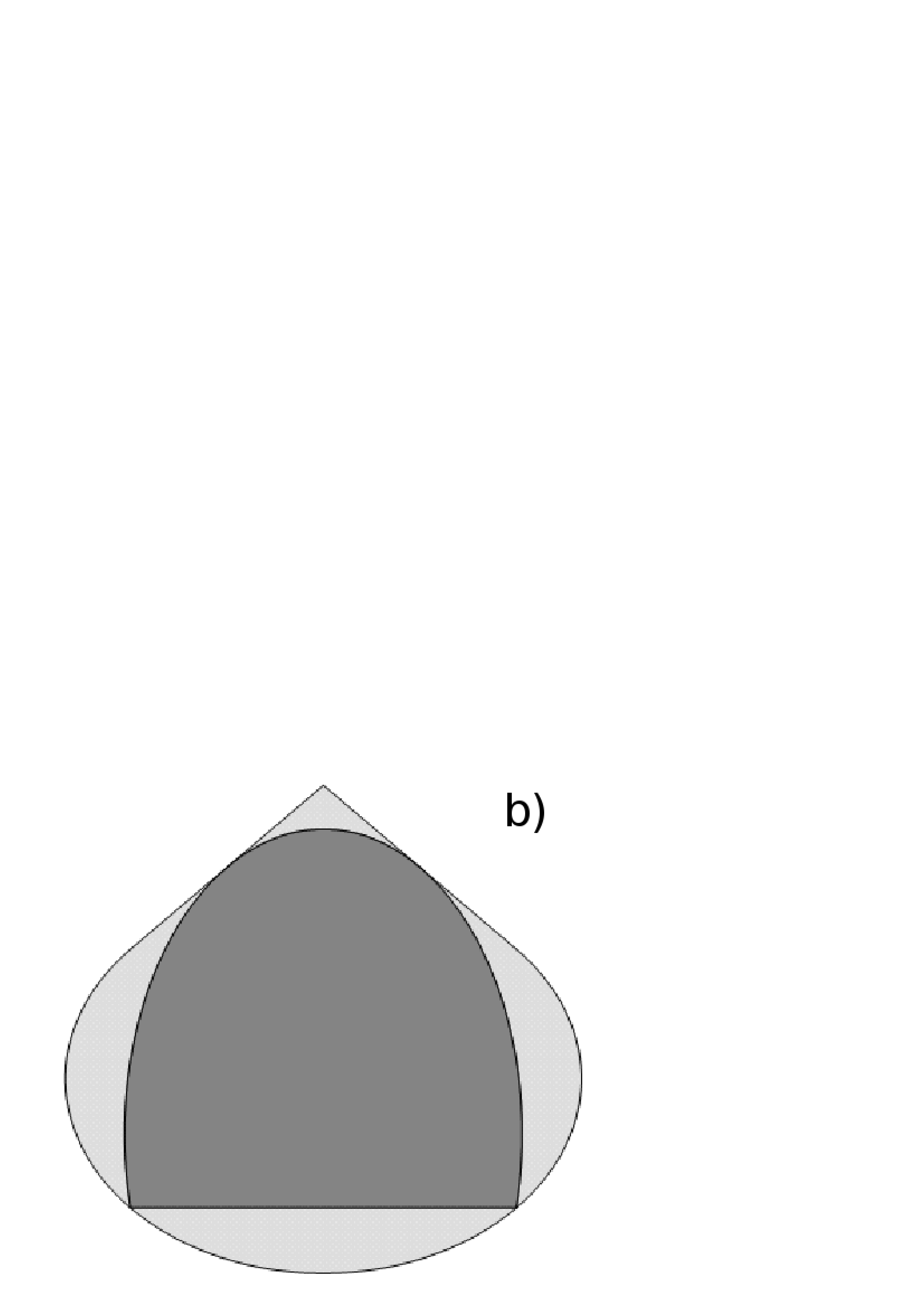}\\
\includegraphics[width=0.20\textwidth,viewport=14 14 398 385,clip=]%
{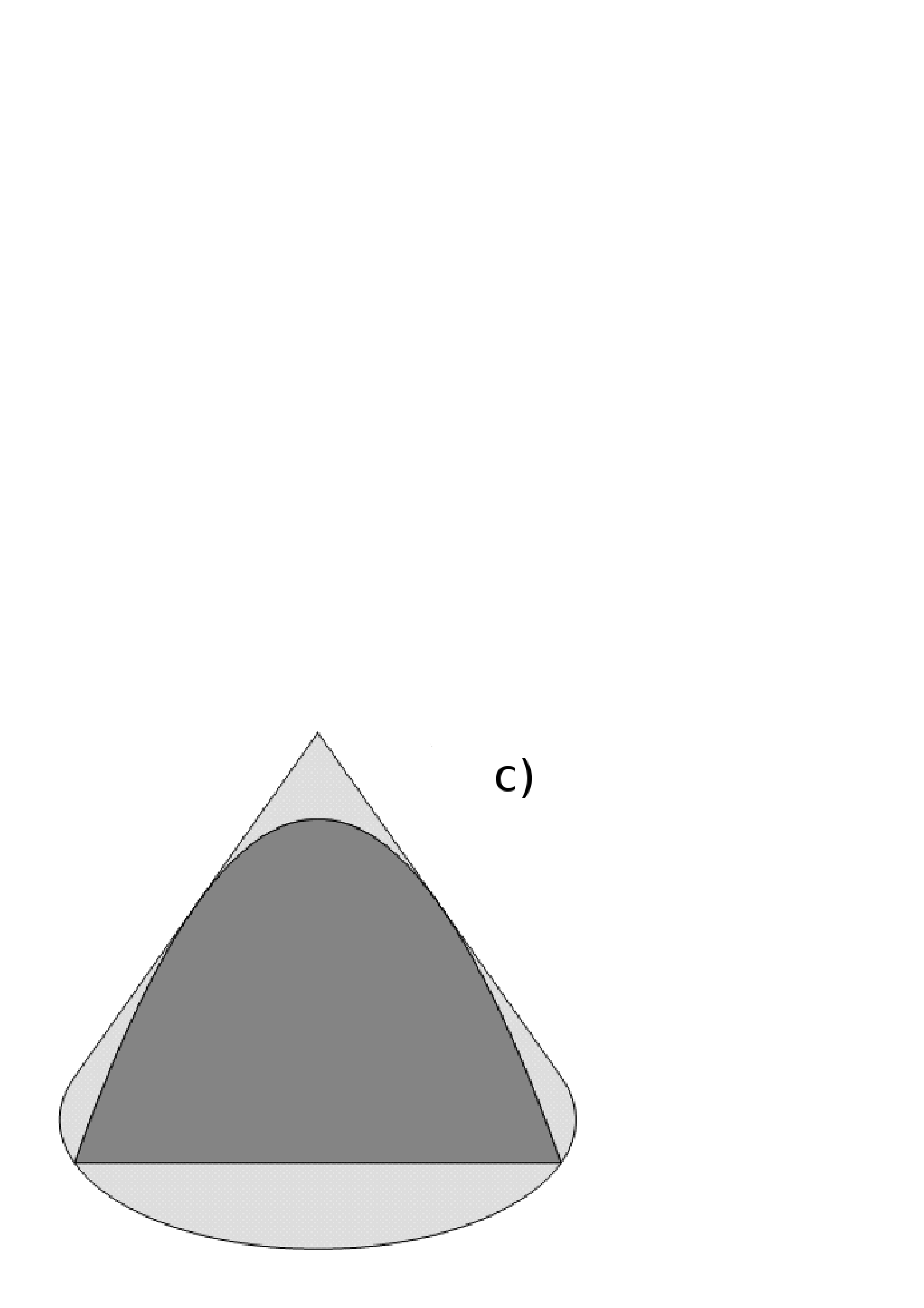}%
\includegraphics[width=0.20\textwidth,viewport=14 14 388 380,clip=]%
{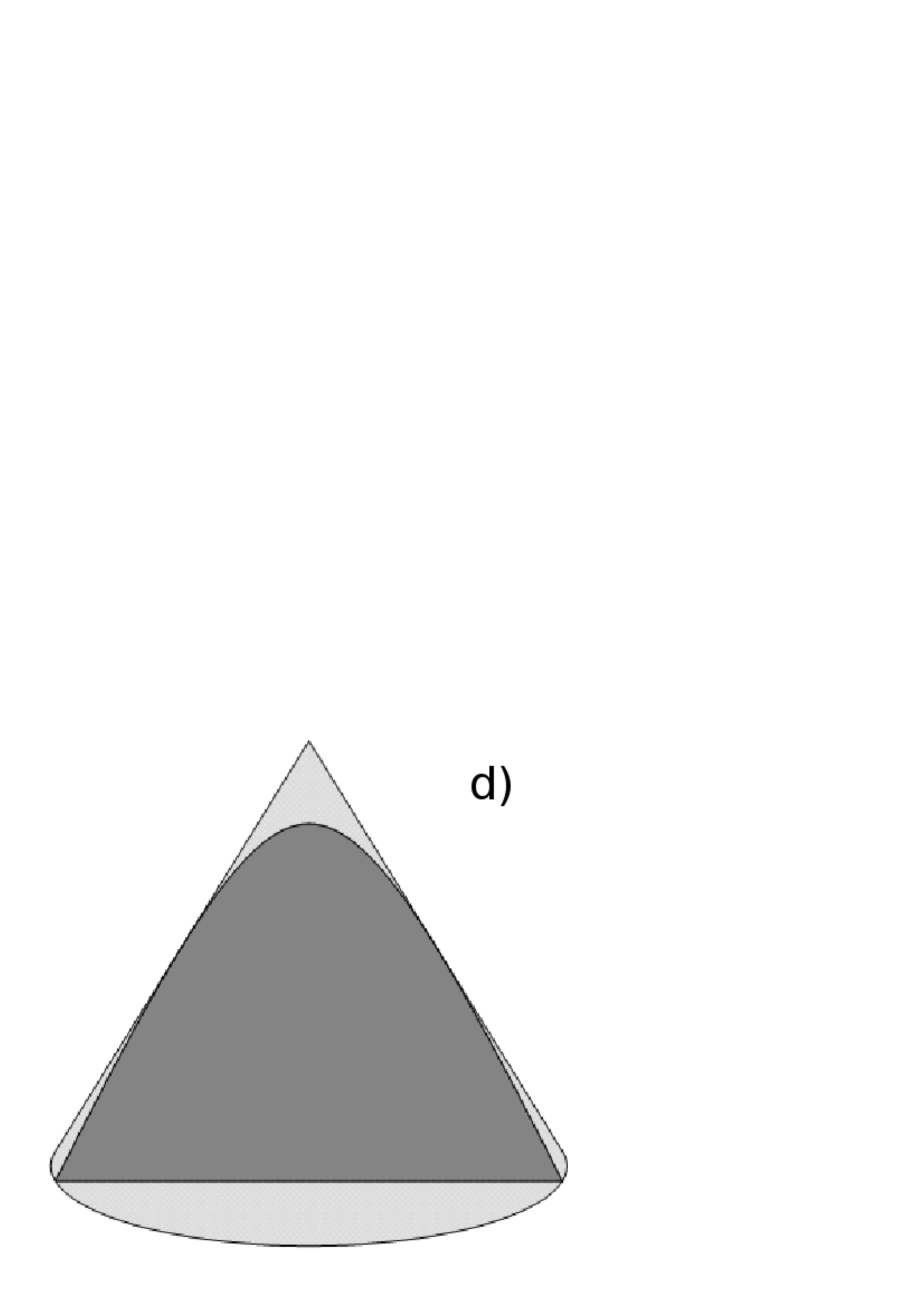}
\end{center}
\caption{\label{fig:2d_pics}
The drawings are dual pairs of planar cross-sections $S_U-\id/3$ (dark)
and projections $P_U$ (bright) of the convex body of qutrit quantum states
${\cal Q}_3$. Drawing a is obtained from the 3D dual pair in 
Fig.~\ref{fig:steiner_cayley} a) and b)--d) are derived from the self-dual
cone in Fig.~\ref{fig:steiner_cayley} b). The cross-sections in b)--d)
have an elliptic, parabolic and hyperbolic boundary piece, respectively.}
\end{figure}

In the case of a matrix $A$ of order $N=3$ the shape of its numerical
range was characterized algebraically in \cite{K51,KRS97}. Regrouping this 
classification we divide the possible shapes into four cases according
to the number of flat boundary parts: The set $W$ is compact and its
boundary $\partial W$ 

\begin{enumerate}
    \item has {\sl no} flat parts. Then $W$ is strictly convex,
    it is bounded by an ellipse or equals the convex hull of a
    (irreducible) sextic space curve;  

    \item has {\sl one} flat part, then $W$ is the convex hull of a
    quartic space curve -- e.g. $W$ is the convex hull of a trigonometric
    curve known as the cardioid;

    \item has {\sl two} flat parts, then $W$ is the convex hull of an
    ellipse and a point outside it;

    \item has {\sl three} flat parts, then $W$ is a triangle
     with corners at eigenvalues of $A$. 
\end{enumerate}

In case 4 the matrix $A$ is normal, $AA^{\dagger}=A^{\dagger}A$, and
the numerical range is a projection of the simplex $\Delta_{2}$ onto a
plane. Looking at the planar projections of ${\cal Q}_3$ shown in 
Fig.\ref{fig:2d_pics} we recognize cases 2 and 3. All four cases are
obtained as projections of the Roman surface in 
Fig. \ref{fig:steiner_cayley} a) or the cone shown in Fig. 
\ref{fig:steiner_cayley} b). A rotund shape and one with two flats
are obtained as a projection of both 3D bodies. A triangle is obtained
from the cone and a shape with one flat from the Roman surface.

In order to actually calculate a $2D$ projection
$P:=\{({\rm Tr}u \rho,{\rm Tr}v \rho)^{\rm T}\in\mathbb R^2
\mid\rho\in {\cal Q}_3\}$ of the set ${\cal Q}_3$ determined by two
traceless hermitian matrices $u$ and $v$ one may proceed  as follows
\cite{HJ2}. For every non-zero matrix $F$ in the real span of $u$ and
$v$ we calculate the maximal eigenvalue $\lambda$ and the corresponding 
normalized eigenvector $|\psi\rangle$ with 
$F|\psi\rangle=\lambda|\psi\rangle$. Then
$(\langle\psi|u|\psi \rangle,\langle \psi |v|\psi \rangle)^{\rm T}$ belongs
to the projection $P$, and these points cover all exposed points of $P$. 

\subsection{Exposed and non-exposed faces}

\begin{figure}
\begin{center}
\includegraphics[width=0.11\textwidth,viewport=14 14 280 367,clip=]%
{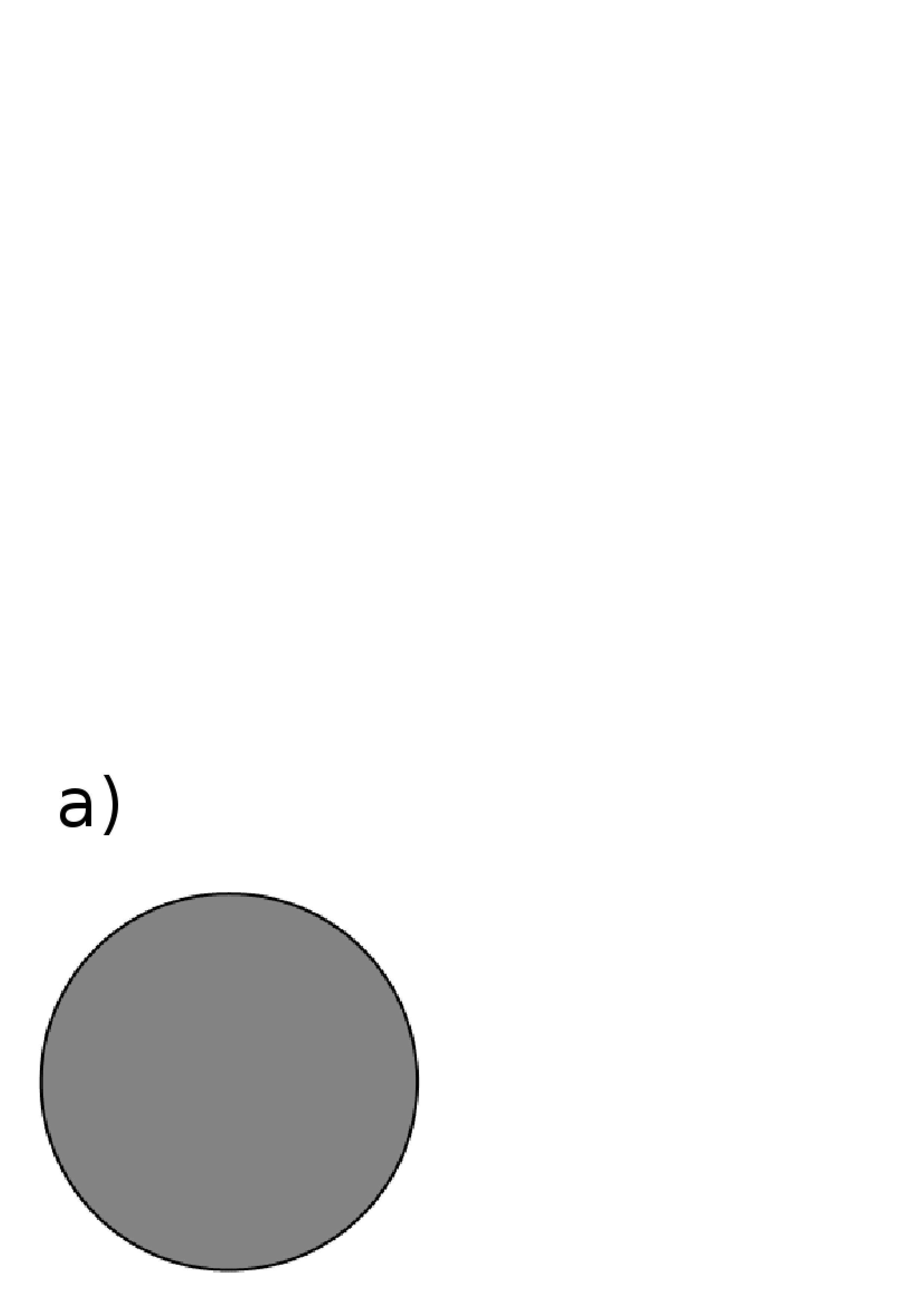}%
\includegraphics[width=0.11\textwidth,viewport=14 14 283 416,clip=]%
{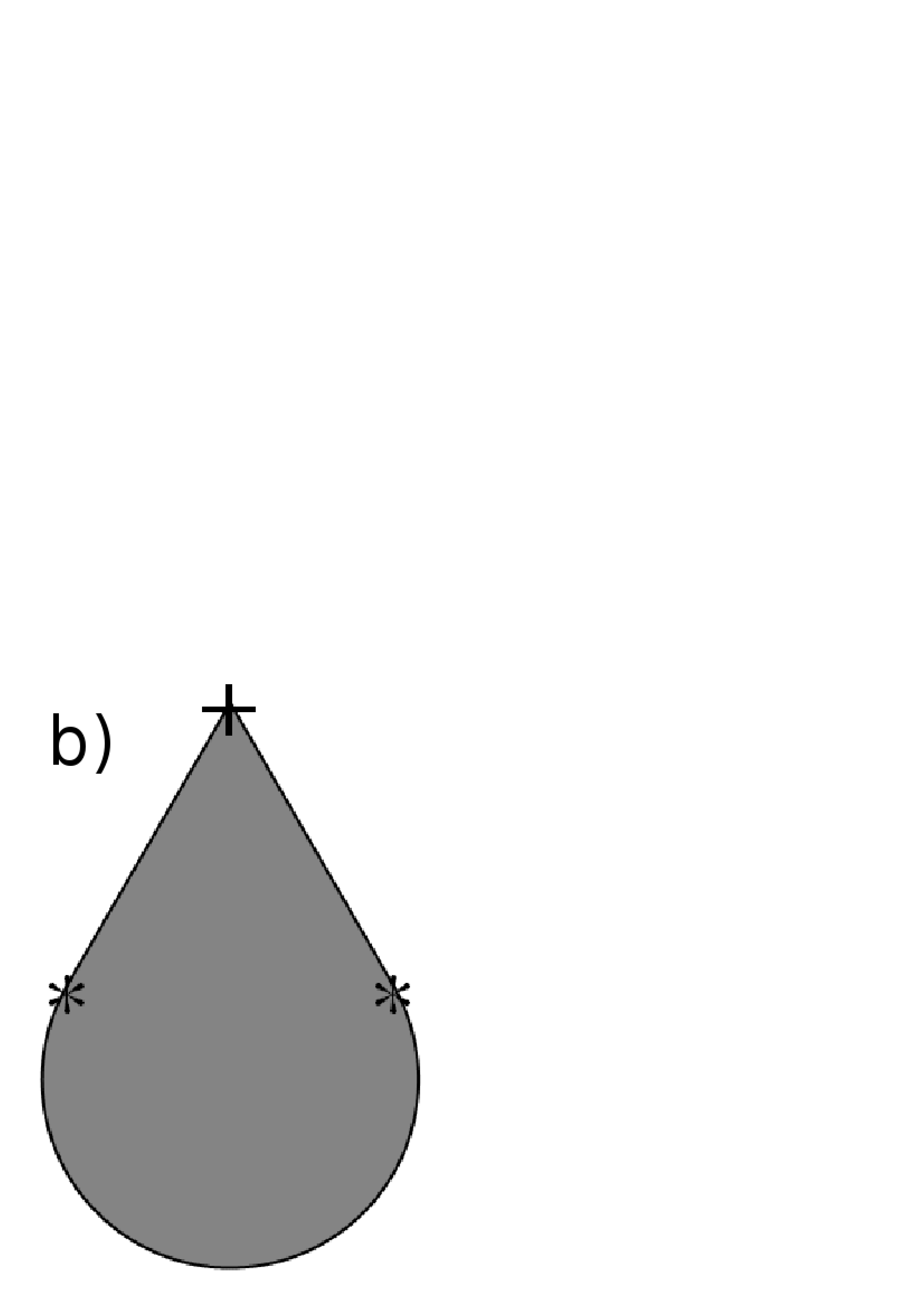}%
\includegraphics[width=0.11\textwidth,viewport=14 14 283 416,clip=]%
{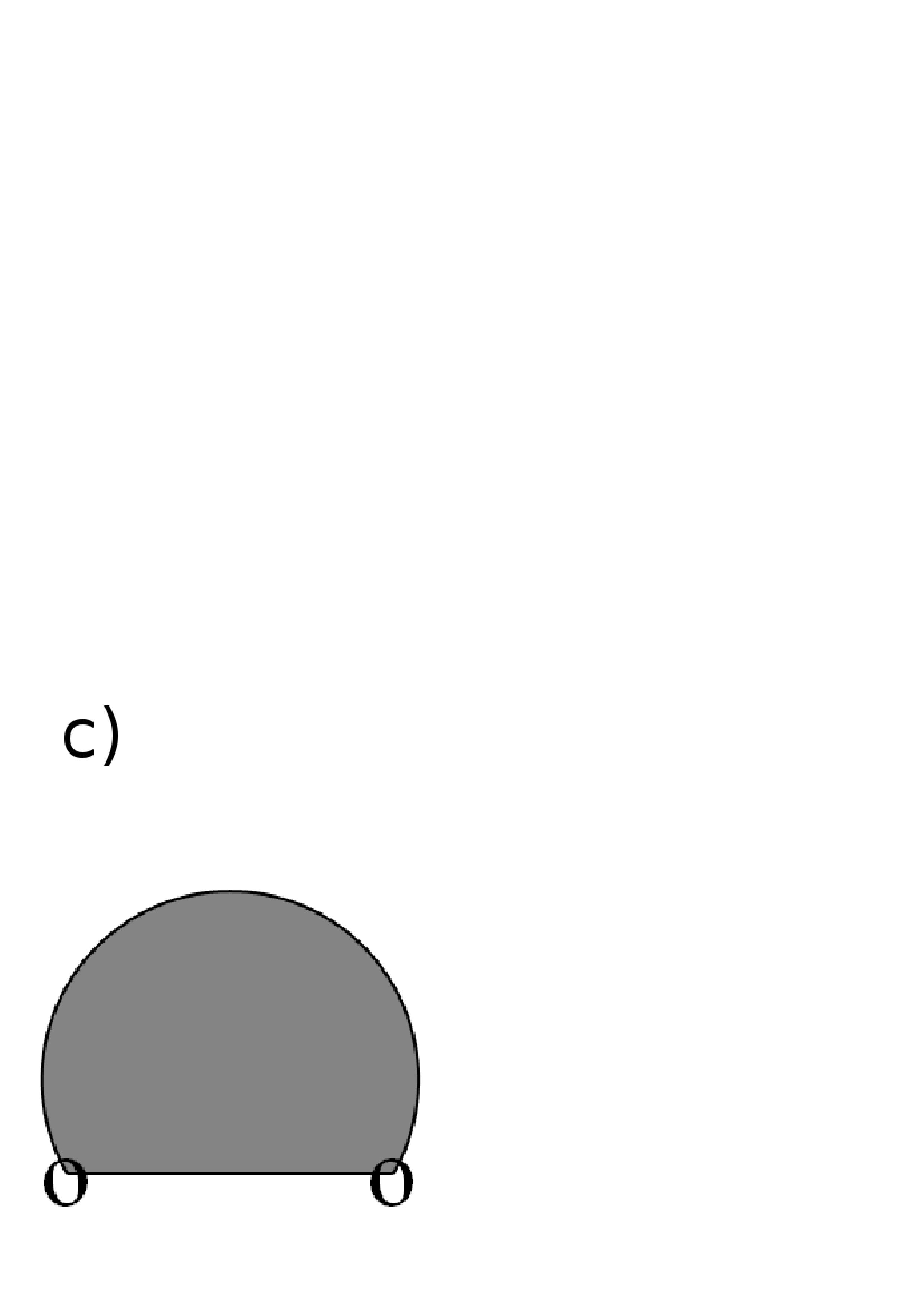}%
\includegraphics[width=0.11\textwidth,viewport=14 14 283 416,clip=]%
{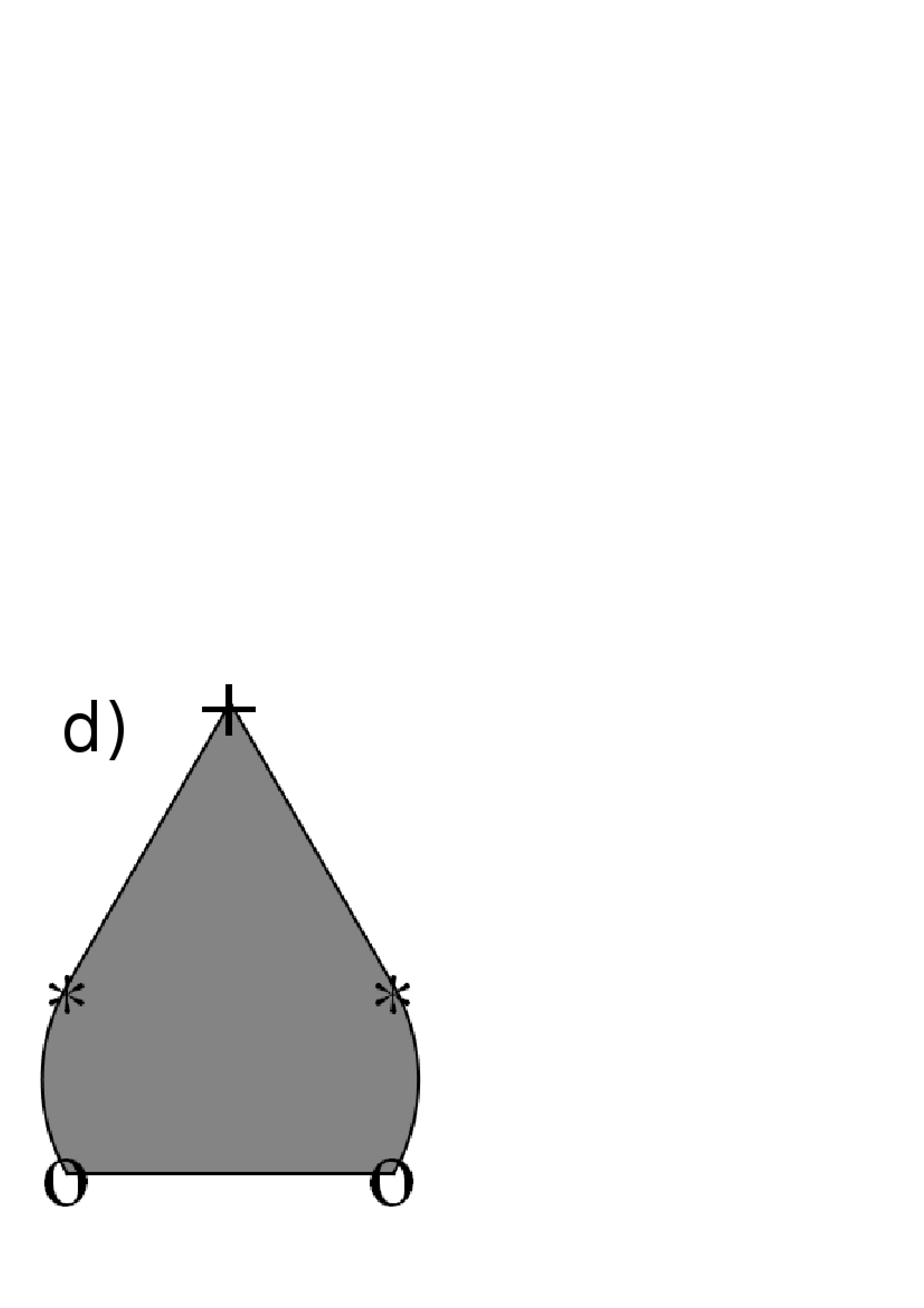}
\end{center}
\begin{tabular}{|c|c|c|c|c|}
\hline \hline
Exemplary sets &  disk $a)$ &   drop $b)$ & 
  \parbox{1.4cm}{  truncated \\ disk $c)$}  & 
  \parbox{1.4cm}{ truncated \\ drop $d)$} \\\hline 
non-exposed points $(*)$ &
no & yes & no & yes\\\hline
\parbox{3.3cm}{\center non-polyhedral
\\ corners  $(o)$} &
no & no & yes & yes \\\hline
set is self-dual & yes & no & no & yes\\
\hline \hline
\end{tabular}
\caption{\label{fig:dual_pics} Exemplary convex sets and their duals.
Symbols: non-exposed point ($*$), polyhedral corners ($+$) and
non-polyhedral corners $(o)$. Sets a) and d) are self-dual, while
b) and c) is a dual pair. Sets a) and c) have properties like 2D
cross-sections of ${\cal Q}_N$, while sets a) and b) could be obtained
from ${\cal Q}_N$ by projection.}
\end{figure}

\medskip

\par
An exposed face of a convex set $X$ is the intersection of $X$ with an 
affine hyperplane $H$ such that $X\setminus H$ is convex, i.e.\ $H$
intersects $X$ only at the boundary. Examples in the plane are the
boundary points of the disk in  Fig.~\ref{fig:dual_pics} a) or the
boundary segments in panels b) and d). A non-exposed face of $X$ is
a face of $X$ that is not an exposed face. In dimension two non-exposed
faces are non-exposed points, they are the endpoints of boundary segment
of $X$ which are not exposed faces by themselves. Examples are the lower 
endpoints of the boundary segments in Fig.~\ref{fig:dual_pics} b) or d). 
\par
It is known that cross-sections of ${\cal Q}_N$ have no
non-exposed faces. On the other hand the twisted cylinder
(see Fig. \ref{fig5}) and the  convex hull $C$ of the space curve 
(Fig.~\ref{fig:hull_curve}) do have non-exposed faces of dimension one. 
In contrast to cross-sections, projections of ${\cal Q}_N$ can 
have non-exposed points, see e.g.\ the planar projections of
${\cal Q}_3$ in Fig.~\ref{fig:2d_pics}. They are related to 
discontinuities in certain entropy functionals (in use as information
measures) \cite{Knauf_Weis}.
\par
The dual concept to exposed face is normal cone \cite{Weis_tc}. 
The normal cone of a two-dimensional convex set $X\subset\mathbb R^2$
at $(x_1,x_2)\in X$ is
\begin{align*}
\{(y_1,y_2)^{\rm T}&\in\bR^2\mid\\ 
& (z_1-x_1)y_1+(z_2-x_2)y_2\leq 0\;\forall
(z_1,z_2)\in X\}.
\end{align*}
The normal cone generalizes outward pointing normal vectors of a 
smooth boundary curve of $X$ to points $(x_1,x_2)$ where this curve is
not smooth. Then the dimension of the normal cone is two and we call
$(x_1,x_2)$ a corner. The examples in Fig.~\ref{fig:dual_pics} have
$0,1,2,3$ corners from left to right. There are different types of
corners: The top corners of Fig.~\ref{fig:dual_pics} b) and d) are
polyhedral, i.e.\ they are intersections of two boundary segments. If 
a corner is not the intersection of two boundary  segments we call it 
non-polyhedral. The bottom corners of c) and d) are non-polyhedral corners. 
Polyhedral and non-polyhedral corners are characterized in \cite{Weis_dn}
in terms of normal cones. From this characterization it follows that any
corner of a two-dimensional projection of ${\cal Q}_N$ is polyhedral 
\cite{Weis_tc}. An analogue property holds in higher dimensions 
but it can not be formulated in terms of polyhedra.
Fig.~\ref{fig:2d_pics} shows that two-dimensional cross-sections of
${\cal Q}_3$ can have non-polyhedral corners.
\par
Given a two-dimensional convex body
including the origin in the interior, the duality (\ref{eq:dual_face})
maps non-exposed points onto the set of non-polyhedral corners of the
dual convex body. There will be one or two non-exposed points in each
fiber depending on whether the corner does or does not lie on a boundary
segment of the dual body \cite{Weis_dn}. We conclude that a two-dimensional 
self-dual convex set has no non-exposed points if and only if all its
corners are polyhedral.

\section{When the dimension matters}
\label{sec_mubsic}

So far we have discussed the qutrit, and properties of the qutrit that 
generalise to any dimension $N$. But what is special about a quantum 
system whose Hilbert space has dimension $N$? The question gains some 
relevance from recent attempts to find direct experimental signatures 
of the dimension, 

One obvious answer is that if and only if $N$ is a composite number, 
the system admits a description in terms of entangled subsystems. But 
we can look for an answer in other directions too. We emphasised that 
a regular simplex $\Delta_{N-1}$ can be inscribed in the quantum state 
space ${\cal Q}_N$. But in the Bloch ball we can clearly inscribe not 
only $\Delta_1$ (a line segment), but also $\Delta_2$ (a triangle) 
and $\Delta_3$ (a tetrahedron). If we insist that the vertices of the 
inscribed simplex should lie on the outsphere of ${\cal Q}_N$, and also 
that the simplex should be centred at the maximally mixed state, then this 
gives rise to a non-trivial problem once the dimension $N > 2$. This 
is clear from our model of the latter as the convex hull of the seam of 
a tennis ball, or in other words because the set of pure states form a 
very small subset of the outsphere. Still we saw, in Fig. 
\ref{fig:steiner_cayley} a), that not only $\Delta_2$ but also $\Delta_3$ 
can be inscribed in ${\cal Q}_3$, and as a matter of fact so can 
$\Delta_5$ and $\Delta_8$. 
But is it always possible to inscribe the regular simplex $\Delta_{N^2-1}$ 
in ${\cal Q}_N$, in such a way that the $N^2$ vertices are pure states? 
Although the answer is not obvious, it 
is perhaps surprising to learn that the answer is 
not known, despite a considerable amount of work in recent years. 

The inscribed regular simplices $\Delta_{N^2-1}$ are known as symmetric 
informationally complete positive operator valued measures, or SIC-POVMs 
for short. Their existence has been established, by explicit construction, 
in all dimensions $N \leq 16$ and in a handful of larger dimensions.
The conjecture is that they always exist \cite{Scott}. But the available 
constructions 
have so far not revealed any pattern allowing one to write down a solution 
for all dimensions $N$. Already here the quantum state space begins to 
show some $N$-dependent individuality.

Another question where the dimension matters concerns complementary 
bases in Hilbert space. As we have seen, given a basis in Hilbert space, 
there is an $(N-1)$-dimensional cross-section of ${\cal Q}_N$ in which 
these vectors appear as the vertices of a regular simplex $\Delta_{N-1}$. 
We can---for instance for tomographic reasons \cite{Wootters}---decide 
to look for two 
such cross-sections placed in such a way that they are totally orthogonal 
with respect to the trace inner product. If the two cross-sections are 
spanned by two regular simplices stemming from two Hilbert space bases 
$\{ |e_i\rangle \}_{i=0}^{N-1}$ and $\{ |f_i\rangle \}_{i=0}^{N-1}$, then 
the requirement on the bases is that 

\begin{equation} |\langle e_i|f_j\rangle |^2 = \frac{1}{N} \end{equation}
for all $i,j$. Such bases are said to be complementary, and form a key 
element in the Copenhagen interpretation of quantum mechanics 
\cite{Schwinger}. But do they exist for all $N$?

The answer is yes. To see this, let one basis be the computational one, 
and let the other be expressed in terms of it as the column vectors of the 
Fourier matrix 

\begin{equation} F_N = \frac{1}{\sqrt{N}}
\left[ \begin{array}{ccccc}1 & 1 & 1 & \dots & 1 \\ 1 & \omega & \omega^2 
& \dots & \omega^N \\ 
1 & \omega^2 & \omega^4 & \dots & \omega^{2(N-1)} \\ 
\vdots & \vdots & \vdots & & \vdots \\ 1 & \omega^{N-1} & \omega^{2(N-1)} 
& \dots & \omega^{(N-1)^2} \end{array} \right] \ , \end{equation}
where $\omega = e^{2\pi i/N}$ is a primitive root of unity. The Fourier matrix 
is an example of a {\sl complex Hadamard matrix}, a 
unitary matrix all of whose matrix elements have the same modulus. 

We are interested in finding all possible complementary pairs up to unitary 
equivalences. The latter are largely fixed by requiring that one member 
of the pair is the computational basis, since the second member will then 
be defined by a complex Hadamard matrix. The remaining freedom is taken 
into account by declaring two complex Hadamard matrices $H$ and $H'$ 
to be equivalent if they can be related by 

\begin{equation} H' = D_1P_1HP_2D_2 \ , \label{Haagerup} \end{equation}
where $D_i$ are diagonal unitary matrices and $P_i$ are permutation matrices.

The task of classifying pairs of cross-sections of ${\cal Q}_N$
forming simplices $\Delta_{N-1}$ and sitting in totally orthogonal
$N$-planes is therefore equivalent
 to the problem of classifying complementary 
pairs of bases in Hilbert space. This problem in turn is equivalent to 
the problem of classifying complex Hadamard matrices of a given size. But 
the latter problem has been open since it was first raised by Sylvester and 
Hadamard, back in the nineteenth century. It has been completely solved only 
for $N \leq 5$, and it was recently almost completely solved for $N = 6$ 
\cite{Szollosi}.

More is known if we restrict ourselves to continuous families of complex Hadamard 
matrices that include the Fourier matrix. Then it has been known for some 
time \cite{TZ08} that the dimension of such a family is bounded from above by 

\begin{equation} d_{F_N} = \sum_{k=0}^{N-1} \mbox{gcd}(k,N) - (2N-1)\ , 
\end{equation}
where gcd denotes the largest common divisor, and gcd$(0,N) = N$. We 
subtracted the $2N-1$ 
dimensions that arise trivially from eq. (\ref{Haagerup}). Moreover, if 
$N = p^k$ is a power of prime number $p$ this bound is saturated by 
families that have been constructed explicitly. In particular, 
if $N$ is a prime number $d_{F_p} = 0$, and the Fourier matrix is an 
isolated solution. For $N = 4$ on the other hand there exists a 
one-parameter family of inequivalent complex Hadamard matrices.

Further results on this question were presented in Bia{\l}owie{\.z}a 
\cite{Nuno}. In 
particular the above bound is not achieved for any $N$ not equal 
to a prime power and not equal to 6. It turns out that the 
answer depends critically on the nature of the prime number decomposition 
of $N$. Thus, if $N$ is a product of two odd primes the answer will look 
different from the case when $N$ is twice an odd prime. However, at the 
moment, the largest non-prime power dimension for which the answer is 
known---even for this restricted form of the problem---is $N = 12$. 
 
At the moment then, both the SIC problem and the problem of complementary 
pairs of bases highlight the fact that the choice of Hilbert space dimension 
$N$ has some dramatic consequences for the geometry of ${\cal Q}_N$. 
Now the basic intuition that drove Mielnik's attempts to generalize quantum 
mechanics was the feeling that the nature of the physical system should 
be reflected in the geometry of its convex body of states \cite{Mi68}. 
Perhaps this intuition will eventually be vindicated within quantum 
mechanics itself, in such a way that the individuality of the system 
is expressed in the choice of $N$?

\section{Concluding remarks}
\label{sec_concl}

Let us try to summarize basic properties of the set ${\cal Q}_N$
of mixed quantum states of size $N\ge 3$
analyzed with respect to the flat, Hilbert-Schmidt
geometry, induced by the distance (\ref{HSdist}).

a) The set ${\cal Q}_{N}$ is a convex set of $N^2-1$ dimensions.
   It is topologically equivalent to a ball
   and does not have  pieces of lower dimensions
   ('no hairs').

b)  The set ${\cal Q}_N$ is inscribed in a sphere of radius 
   $R_N=\sqrt{(N-1)/2N}$, and it contains the maximal ball of radius 
    $r_N=1/\sqrt{2N(N-1)}$.

c) The set  ${\cal Q}_N$  is neither a polytope nor a smooth body.

d) The set of mixed states is self-dual (\ref{eq:dualmN}).

e) All cross-sections of ${\cal Q}_N$ have no
   non-exposed faces.
 
f) All corners of two-dimensional projetions of ${\cal Q}_N$ are
   polyhedral.

g)  The boundary $\partial {\cal Q}_N$ contains
       all states of less than maximal rank.

h) The set of extremal (pure) states forms a connected $2N-2$ dimensional 
   set, which has zero measure with respect to the $N^2-2$ dimensional
   boundary  $\partial {\cal Q}_N$.

i) Explicit formulae for the volume $V$ and the area $A$ of the 
   $d=N^2-1$ dimensional set ${\cal Q}_N$ are known \cite{ZS03}.
   The ratio $Ar/V$ is equal to the dimension $d$, which implies that
   ${\cal Q}_N$ has a constant height \cite{SBZ06} and can be decomposed 
   into pyramids of equal height having all their apices at the centre
   of the inscribed sphere.

\medskip

It is a pleasure to thank Marek Ku{\'s} and Gniewomir Sarbicki 
for fruitful discussions and helpful remarks.
I.B. and K.{\.Z}. are thankful for an invitation for the workshop to
Bia{\l}owie{\.z}a,  where this work was presented and improved.
Financial support by the grant number  N N202 090239 of
Polish Ministry of Science and Higher Education 
and by the Swedish Research Council under contract 
VR 621-2010-4060 is gratefully acknowledged.

\appendix
\section{Trigonometric curves}
\label{app:C}

\par
We write the convex hull $C$ of the trigonometric space curve in 
Section~\ref{sec3} as a projection of a cross-section of the 
$35$-dimensional set ${\mathcal Q}_6$ of density matrices. Up to the
trace normalization, this problem is solved in \cite{Henrion} for
the convex hull of any trigonometric curve $[0,2\pi)\to\mathbb R^n$.
The assumptions are that each of the $n$ coefficient functions
of the curve is a trigonometric polynomial of some finite degree $2d$,
\[\textstyle
t \;\mapsto\;
\sum_{k=1}^d(\alpha_k\cos(kt)+\beta_k\sin(kt))+\gamma
\]
for real coefficients $\alpha_k,\beta_k,\gamma$.
\par
The space curve (\ref{curvec}) lives in dimension $n=3$, we denote its
coefficients by ${\vec x}=(x_1,x_2,x_3)^{\rm  T}$. Using
trigonometric formulas and the parametrization 
$\cos(t)=\tfrac{y_0^2-y_1^2}{y_0^2+y_1^2}$ and
$\sin(t)=\tfrac{2y_0 y_1}{y_0^2+y_1^2}$ we have
\[\begin{array}{rcl}
1 & \stackrel{\text{def.}}{=} & (y_0^2+y_1^2)^4 \ ,\\
x_1 & = &
(y_0^2-y_1^2)^2[(y_0^2-y_1^2)^2-3(2y_0y_1)^2] \ ,\\
x_2 & = &
(y_0^2-y_1^2)(2y_0y_1)[3(y_0^2-y_1^2)^2-(2y_0y_1)^2] \ ,\\
x_3 & = & -(y_0^2+y_1^2)^3(2y_0y_1) \ .
\end{array}\]
A basis vector of $m$-variate forms of degree $2d=8$ is given by
$\vec \xi=(x_0^8,x_0^7x_1,x_0^6x_1^2,x_0^5x_1^3,x_0^4x_1^4,x_0^3x_1^5,
x_0^2x_1^6,x_0x_1^7,x_1^8)$ (for the number $m=1$ used in \cite{Henrion}
for the degrees of freedom of the projective coordinates $(y_0:y_1)$ in
the circle $\mathbb P^1(\mathbb R)$) and we have
\[
(1,x_1,x_2,x_3)^{\rm  T}
\;=\;
A\vec \xi
\]
for the $4\times 9$-matrix
\[
A\;=\;
\left(\begin{smallmatrix}
1 & 0 & 4 & 0 & 6 & 0 & 4 & 0 & 1\\
1 & 0 & -16 & 0 & 30 & 0 & -16 & 0 & 1\\
0 & 6 & 0 & -26 & 0 & 26 & 0 & -6 & 0\\
-1 & 0 & -2 & 0 & 0 & 0 & 2 & 0 & 1
\end{smallmatrix}\right) \ .
\]
Let us denote by $M\succeq 0$ that a complex square matrix $M$
is positive semi-definite. The $5\times 5$ moment matrix of
$\vec u=(u_1,\ldots,u_9)$ is given by
\[
M_4(\vec u)
\;=\;
\left(\begin{smallmatrix}
u_1 & u_2 & u_3 & u_4 & u_5 \\
u_2 & u_3 & u_4 & u_5 & u_6 \\
u_3 & u_4 & u_5 & u_6 & u_7 \\
u_4 & u_5 & u_6 & u_7 & u_8 \\
u_5 & u_6 & u_7 & u_8 & u_9 
\end{smallmatrix}\right) \ .
\]
Now \cite{Henrion} provides the convex hull representation 
\begin{align}
\label{eq:henrion}
& C 
\;\stackrel{\text{def.}}{=}\;
{\rm conv}\{\vec x(t)\in\mathbb R^3\mid t\in[0,2\pi)\}
\;=\\
& \{\left(\begin{smallmatrix}v_1\\v_2\\v_3\end{smallmatrix}\right)
\in\mathbb R^3\mid\exists\vec u\in\mathbb R^9\text{ s.t. }
\left(\begin{smallmatrix}1\\v_1\\v_2\\v_3\end{smallmatrix}\right)
=A\vec u \text{ and } M_4(\vec u)\succeq 0 \}\nonumber
\end{align}
which we shall simplify by eliminating the variables $u_1,\ldots,u_4$. 
\par
A particular solution of $(1,v_1,v_2,v_3)^{\rm T}=A\vec u$ is 
\begin{align*}
\widetilde{u}_1\ =\ \tfrac 1{5}(4+v_1)\ ,\quad
& \widetilde{u}_2\ =\ \tfrac 1{44}(3v_2-13v_3)\ ,\\
\widetilde{u}_3\ =\ \tfrac 1{20}(1-v_1)\ ,\quad
& \widetilde{u}_4\ =\ \tfrac 1{44}(-v_2-3v_3)\ ,
\end{align*}
$\widetilde{u}_5=\widetilde{u}_6=\widetilde{u}_7=\widetilde{u}_8
=\widetilde{u}_9=0$. The reduced row echelon form of $A$ being
\[
\left(\begin{smallmatrix}
1 & 0 & 0 & 0 & 54/5 & 0 & 0 & 0 & 1 \\
0 & 1 & 0 & 0 & 0 & 39/11 & 0 & 2/11 & 0 \\
0 & 0 & 1 & 0 & -6/5 & 0 & 1 & 0 & 0 \\
0 & 0 & 0 & 1 & 0 & -2/11 & 0 & 3/11 & 0
\end{smallmatrix}\right) 
\]
and regarding $u_5,\ldots,u_9$ as free variables we have
\begin{align*}
& u_1\ =\ \widetilde{u}_1 - \tfrac{54}{5}u_5-u_9\ ,
& u_2\ =\ \widetilde{u}_2\ - \tfrac{39}{11}u_6 - \tfrac{2}{11}u_8\ ,\\
& u_3\ =\ \widetilde{u}_3 + \tfrac 6{5}u_5-u_7\ ,
& u_4\ =\ \widetilde{u}_4 + \tfrac 2{11}u_6 - \tfrac 3{11}u_8\ .
\end{align*}
\par
One problem remains, the matrix $M_4$ parametrized by $v_1,v_2,v_3$
and $u_5,\ldots,u_9$ has not trace one,
\[
{\rm Tr}M_4
\;=\;
u_1+u_3+u_5+u_7+u_9
\;=\;
\tfrac 1{20} (17 - 172 u_5 + 3 v_1) \ .
\]
This we correct by adding a direct summand to $M_4$ and by defining
\[
M
\;=\;
\left(\begin{array}{c|c}
M_4 & 0 \\\hline
0 & \tfrac{172}{20} u_5 + \tfrac 3{20} (1 - v_1)
\end{array}\right)\ .
\]
If $M_4\succeq 0$ then $u_5\geq 0$ follows because $u_5$ is a diagonal 
element of $M_4$ and $-1\leq v_1\leq 1$ follows from (\ref{eq:henrion}) 
because $(v_1,v_2,v_3)\in C$ is included in the unit ball of 
$\mathbb R^3$. This proves $M\succeq 0\iff M_4\succeq 0$ and we get
\[ 
C 
\;=\;
\{\left(\begin{smallmatrix}v_1\\v_2\\v_3\end{smallmatrix}\right)
\in\mathbb R^3\mid\exists
\left(\begin{smallmatrix}u_5\\\vdots\\u_9\end{smallmatrix}\right)
\in\mathbb R^5\text{ s.t. }
M\succeq 0 \}\ .
\]
We conclude that $C$ is a projection of the $8$-dimensional 
spectrahedron
$\{(v_1,v_2,v_2,u_5,\ldots,u_9)\in\bR^{3+5}\mid M\succeq 0\}$, which
is a cross-section of $\mathcal Q_6$.

\end{document}